\documentclass[aps,prx,twocolumn,amsmath,amssymb,superscriptaddress,showpacs]{revtex4-2}
\usepackage{graphicx}
\usepackage{dcolumn}
\usepackage{bm}
\usepackage{hyperref}
\usepackage{xcolor}
\usepackage{float} 
\usepackage{soul}
\usepackage{siunitx}

\begin{document}
\title{\textbf{ 
    Elastic interactions compete with persistent cell motility to drive durotaxis }}

\author{Subhaya Bose}
\thanks{These two authors contributed equally}
\affiliation{Department of Physics, University of California Merced, CA 95343, USA}

\author{Haiqin Wang} 
\thanks{These two authors contributed equally}
\affiliation{Technion -- Israel Institute of Technology, Haifa 3200003, Israel}
\affiliation{Department of Physics and MATEC Key Lab, Guangdong Technion - Israel Institute of Technology, 241 Daxue Road, Shantou, Guangdong 515063, China}

\author{Xinpeng Xu}
\thanks{xu.xinpeng@gtit.edu.cn}
\affiliation{Department of Physics and MATEC Key Lab, Guangdong Technion - Israel Institute of Technology, 241 Daxue Road, Shantou, Guangdong 515063, China}
\affiliation{Technion -- Israel Institute of Technology, Haifa 3200003, Israel}

\author{Arvind Gopinath}
\thanks{agopinath@ucmerced.edu}
\affiliation{Department of Bioengineering, University of California Merced, CA 95343, USA
}

\author{Kinjal Dasbiswas}
\thanks{kdasbiswas@ucmerced.edu}
\affiliation{Department of Physics, University of California Merced, CA 95343, USA}

\date{\today}

\begin{abstract}
Many animal cells that crawl on extracellular substrates exhibit durotaxis, \textit{i.e.} directed migration towards stiffer substrate regions.  This has implications in several biological processes including tissue development and tumor progression.  Here, we introduce  a phenomenological model for single cell durotaxis that incorporates both elastic deformation-mediated cell-substrate interactions and the stochasticity of cell migration. Our model is motivated by a key observation in an early demonstration of durotaxis: a single, contractile cell at a sharp interface between a softer and a stiffer region of an elastic substrate reorients and migrates towards the stiffer region.  We model migrating cells as self-propelling, persistently motile agents that exert contractile traction forces on  their elastic substrate. The resulting substrate deformations induce elastic interactions with mechanical boundaries, captured by an elastic potential. The dynamics is determined by two crucial parameters: the strength of the cellular traction-induced boundary elastic interaction ($A$), and the persistence of cell motility ($Pe$).  Elastic forces and torques resulting from the potential orient cells perpendicular (parallel) to the boundary and accumulate (deplete) them at the clamped (free) boundary. Thus, a clamped boundary induces an attractive potential that drives durotaxis, while a free boundary induces a repulsive potential that prevents anti-durotaxis. By quantifying the steady state position and orientation probability densities, we show how the extent of accumulation (depletion) depends on the strength of the elastic potential and motility. We compare and contrast crawling cells with biological microswimmers and other synthetic active particles, where accumulation at confining boundaries is well-known. We define metrics quantifying boundary accumulation and durotaxis, and present a phase diagram that identifies three possible regimes: durotaxis, and adurotaxis with and without motility-induced accumulation at the boundary. Overall, our model predicts how durotaxis depends on cell contractility and motility, successfully explains some previous observations, and provides testable predictions to guide future experiments.

\begin{description}
\item[Keywords]
\small{Mechanobiology, Cell Motility, Elastic dipoles, Active Brownian Particles, Durotaxis} 
\end{description}
\end{abstract}
\newcommand{\kd}{\textcolor{blue}}
\newcommand{\haiqin}{\textcolor{cyan}}
\newcommand{\sub}{\textcolor{brown}}
\newcommand{\vect}[1]{\boldsymbol{#1}}
\maketitle
%
%
\section{Introduction}
Animal cells migrate by crawling on elastic substrates during many crucial biological processes such as wound healing, tumor progression and tissue development \cite{yamada2019mechanisms}.
Cell migration is responsive to physical cues of their extracellular environment, such as extent and degree of confinement and stiffness of the ambient material or the substrate \cite{charras2014physical}. 
Migrating cells consume energy in the form of ATP to generate directed motion interspersed with stochastic reorientations.  Cell trajectories may thus be represented by active particle models \cite{Romanczuk2012}, where ``active'' implies autonomous energy-consuming units that generate their own motion. Collections of such active particles constitute out-of-equilibrium complex systems and exhibit unusual statistical 
properties such as motility-induced phase separation and accumulation at confining boundaries \cite{redner2013structure, Cates2015}. The extracellular matrix of migrating cells are typically heterogeneous in stiffness and geometry, which implies that cell migration is influenced by mechanical boundaries. Motile cells may therefore be considered as living active matter that interact with their complex environments \cite{Bechinger2016}. 

While crawling cells may exhibit different migration modes \cite{sengupta2021principles}, they share common mechanical processes underlying their motion.  Migration relies on the formation of actin polymerization-induced protrusions at the leading edge,  myosin-motor induced retraction at the trailing edge, adhesive interactions at the cell-substrate interface, \cite{alberts2017molecular} as well as dynamic positioning of the cell nucleus \cite{friedl2011nuclear}. 
These components are coupled by the polarizable active cytoskeleton and together play the dual role of sensing the cell's local microenvironment and driving its net motion. At the cellular scale, this machinery leads to coordinated, directed migration, which
manifests as persistent motion interspersed with speed and orientation fluctuations on uniform two-dimensional substrates. The complex polarity processes and protrusion formation can be effectively captured by the self-propulsion speed with a characteristic persistent time scale, and the translational noise in phenomenological models for cell motility \cite{selmeczi2005cell}. As cells migrate, they exert traction forces on the underlying substrate.  These forces are generated within the cell by its actomyosin cytoskeletal machinery and are communicated to the extracellular substrate through localized focal adhesions \cite{gardel_10}. These traction forces can be significant and generate measurable deformation in the elastic extracellular substrate \cite{pelham_97,Schwarz2001}. 

By actively deforming the substrate, cells sense geometric and mechanical cues in their micro-environment, including material properties such as the substrate stiffness \cite{Sam2013a} and viscoelasticity \cite{Chaudhuri2020}. This gives rise to the possibility of long range cell-cell mechanical communication mediated by mutual deformations of the elastic substrate \cite{Bischofs2006}, for which there is mounting experimental evidence.  For example, endothelial cells modulate their pairwise inter-cellular contact frequency according to substrate stiffness \cite{Reinhart2008}, while forming multicellular networks on substrates of appropriate stiffness \cite{ Rudiger2020, Califano2008}. Recent theoretical works show that both these trends can be quantitatively understood through substrate-mediated cell-cell mechanical interactions \cite{Bose2021, noerr2023optimal}.
The cells may use these mechanical cues, in addition to chemical signaling that is ubiquitous in biology, to direct their persistent migration \cite{Angelini2010, notbohm2016cellular}. 

The observed preferential migration of cells along gradients in substrate stiffness, usually towards stiffer regions, has been termed ``durotaxis'' \cite{lo2000cell, shellard2021durotaxis, sunyer2020durotaxis}. Durotaxis has been observed both in  single cells in culture \cite{raab2012crawling, yeoman2021adhesion, Park2016, duchez2019durotaxis}, as well as in collections of confluent migrating cells \cite{sunyer2016collective}, including \textit{in vivo} \cite{shellard2021collective}. Small cell clusters have also been observed to exhibit negative durotaxis and migrate towards softer substrate regions \cite{isomursu2022directed}. Durotaxis is influenced by matrix composition, as observed in the case of vascular smooth muscle cells on fibronectin substrates but not on cells on laminin-coated substrates \cite{Hartman2016}.  
Suggested biophysical mechanisms for durotaxis include enhanced persistent cell motility due to enhanced cell polarization on stiffer substrates \cite{yu2017phenomenological}, larger local deformation of the softer substrate when the cell or collective is spread across a gradient resulting in overall translation of the center of mass towards the stiffer side \cite{sunyer2016collective, isomursu2022directed}, and more stable focal adhesions on the stiffer side. While the higher persistence of cell motion on stiffer substrates may be rationalized based on the strongly polarized cell shapes in stiffer environments \cite{zemel_10, novikova2017persistence}, this does not address the important roles of cell traction forces exerted on the substrate, and of cell-substrate adhesion, in driving durotaxis.  Recent work using molecular clutch models at the level of single cells or confluent tissue have explained  durotaxis as arising from  stiffness-dependent cell-substrate adhesive interaction \cite{sunyer2016collective, isomursu2022directed, chen2022, shu2023effects}. However, these mechanistic models do not lend easily to the evaluation of the statistical distributions of numerous cell trajectories at long times.

Experimental measurements of time-averaged traction forces mapped to cell shapes \cite{Mandal2014} suggest that stresses can be effectively resolved into a contractile force dipole acting along a preferred axis \cite{gardel_15}. Thus, traction force patterns exerted by a cell on underlying elastic substrates may be modeled as a force dipole. This force distribution also satisfies internal force balance \cite{Sam2013a} as required.  Such a minimal theoretical description of traction forces exerted by an adherent cell leads to a natural organization principle for cells in compliant media \cite{Bischofs2003}. By orienting along directions of maximal stretch, as well as moving towards stretched regions of the substrate, a contractile cellular force dipole can lower the elastic deformation energy of the substrate. This naturally leads to configuration-dependent torques and forces that may drive directed motion  or durotaxis of the cellular force dipole near an elastic interface between a softer and stiffer region \cite{Bischofs2004}. While this static theoretical model predicts the alignment and attraction of the cell towards the stiffer region, it does not address how a self-propelling cell with intrinsically noisy dynamics moves to this favored configuration.

\begin{figure*}[hbt!]
\centering
\includegraphics[width=0.81\linewidth]
{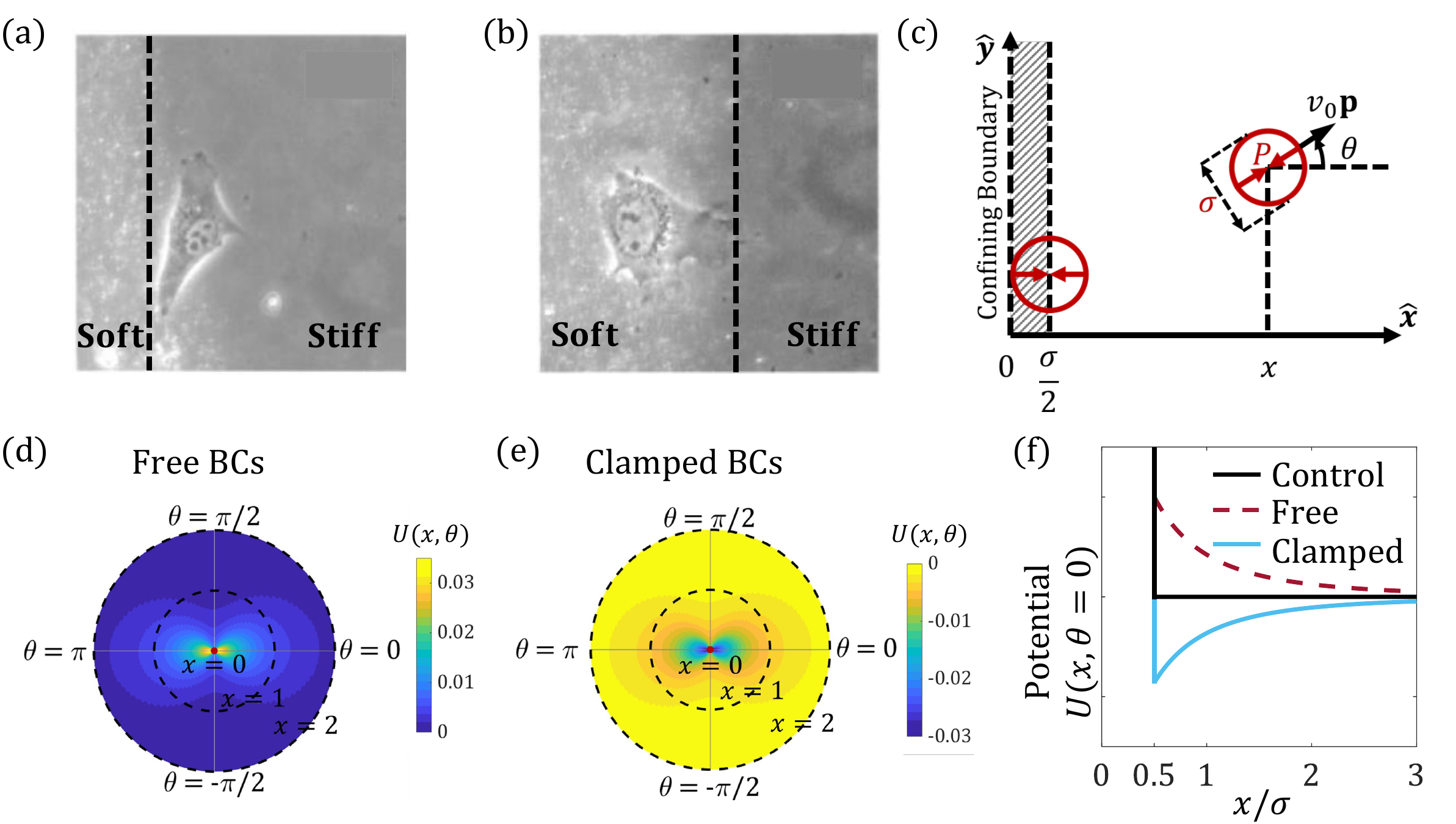}
\caption{
\textbf{Experimental motivation and model setup.}
(a, b) Isolated fibroblasts near interfaces between soft and stiffer regions of a polyacrylamide gel substrate (reproduced with permission from Ref.~\cite{lo2000cell}). (a) A cell approaching the interface from the stiffer side (left) aligns parallel to the interface and remains in the stiffer region. 
(b) A cell on the softer side aligns normal to the interface and eventually crosses over to the stiffer side. 
(c) Schematic of a cell, modeled as circular disc of diameter $\sigma$, moving on a flat linear elastic substrate with uniform stiffness (given by Young's modulus 
$E$, and Poisson's ratio $\nu$) near a confining boundary.  Clamped or free elastic boundary conditions are used to model the cell being on the  
softer or stiffer region of the substrate, respectively. 
Unlike in experiment, the simulated cell is confined and not allowed to cross the boundary. Traction forces generated by the cell are reduced to a contractile force dipole of strength ${\mathbf{P}}$ (red, inward pointing arrows) acting on the substrate. 
The direction of propulsion ${\mathbf p}$ is assumed to be along the cell dipole axis and makes an angle $\theta$ with the horizontal axis.  The cell lies a horizontal distance $x$ from the boundary (the $y$ axis). An excluded region of extent $\sigma/2$ (a lower limit) at the boundary models confinement. (d,e) The spatial map of the elastic interaction potential experienced by the cell as a function of distance from the boundary and its orientation is shown for free and clamped boundaries, respectively. (f) The elastic potential is shown as a function of distance for the control case representing pure confinement without elastic interactions (solid black), the repulsive free boundary (dashed, brown) and the attractive clamped boundary (solid, cyan).}   
\label{fig:figure_schematic}
\end{figure*}

A complete description of durotaxis thus requires combining the elastic model for cell traction-induced matrix deformations by adherent cells, with an appropriate model for stochastic cell movement \cite{Palmieri2019, Bose2021, godeau20223d}. We consider here persistently motile cells that move in a directed manner for a characteristic time before reorienting. Since migrating cells generate protrusions that may be randomly driven by noisy internal signalling \cite{sens2020stick}, the motion of our model cells feature stochastic reorientations and velocity fluctuations \cite{selmeczi2005cell}. Cells are assumed to  
move persistently and exert traction along their long axis, such that the the polarization coincides with their principal traction axis  \cite{carlsson2011mechanisms, Ron2020}. 
We here propose and study a general, phenomenological 
model that incorporates these key elements to provide a statistical physics description of durotaxis. 

Figs.~\ref{fig:figure_schematic}(a) and (b), reproduced from Ref.~\cite{lo2000cell}, illustrate the scenario we wish to analyze theoretically. The authors here examined the behavior of a  fibroblast cell cultured on a deformable polyacrylamyde hydrogel substrate, and located near an interface separating a soft region from a stiffer region. When the cell is on the stiff side, it aligns parallel to the interface and remains on the stiffer side. On the other hand, when the cell starts off on the soft side, it aligns perpendicular to the interface and eventually moves and crosses over to the stiffer side (not shown). This behavior may be understood by considering the polarized cell as a force dipole acting along its axis of elongation \cite{Bischofs2004}. When on the stiffer side (Fig. \ref{fig:figure_schematic}(a)), the cell deforms the interface and the softer elastic medium on the other side of the interface can easily displace, resulting in an effectively stress-free boundary condition. Conversely, when the cell is on the soft side (Fig. \ref{fig:figure_schematic} (b)), the rigid medium on the other side undergoes minimal displacement at the interface, resulting in an effectively clamped boundary. 
In fact, it was shown in Ref.~\cite{Bischofs2004}  that when the interface acts as a clamped (free) boundary, the effective elastic interaction potential between a cell dipole and the interface computed by a full consideration of the virtual image stress distribution required to satisfy the relevant boundary condition, yields an attractive (repulsive) force on the dipole. Additionally, elastic interactions also result in a torque that orients the dipole perpendicular (parallel) to the interface.

While the static model for an adherent cell provides a heuristic explanation for single-cell durotaxis ~\cite{Bischofs2004}, we consider here the role of cell motility, in the presence of an elastic boundary interaction arising from cell traction.
Unlike the original durotactic experiment ~\cite{lo2000cell}, we also choose to confine the model cell to either the softer or stiffer region. This mimics complex or micro-patterned environments and allows us to study the interplay of motility,confinement and elastic interactions.  The model setup of a cell moving on an elastic substrate near a confining boundary is illustrated in Fig.~\ref{fig:figure_schematic}c.
The substrate deformation-mediated elastic interaction potential experienced by a stationary cell is depicted in Figs.~\ref{fig:figure_schematic} (d)-(f). The elastic potential as a function of the cell orientation is shown for free and clamped boundaries in Figs.~\ref{fig:figure_schematic}(d) and (e), respectively that highlight the repulsive and attractive nature of the interactions, as well as the favored parallel and perpendicular cell orientations. Fig.~\ref{fig:figure_schematic}(f) shows the long-range spatial decay of the potential away from the interface in the clamped, free and ``control'' regions, the last corresponding to only steric interactions with the confining boundary. Using this model, described in more detail in the next section, we seek to predict how statistical distributions of cells depend on the persistent and stochastic aspects of motility, as well as the strength and nature of the elastic interactions with the boundary.

 \section{Model for cell motility and elastic cell-boundary interactions}
The   motion of each cell is modeled using Langevin dynamics in the overdamped limit since inertial effects are negligible at the microscale. Each cell is treated as a disk of diameter of diameter $\sigma$ moving on a 2D $xy$-plane corresponding to the surface of an idealized, infinitely thick elastic substrate. 
The state of each cell is defined by its position vector ${\bf r}$ corresponding to the cell center, and unit orientation vector ${\bf p}$ associated with its self-propulsion direction (Fig. 1(c)). Cells move with speed $v_{0}$ in the direction $\mathrm{\bf p}$ (with Cartesian components $(\cos\theta, \sin\theta)$), and interact with boundaries through a potential $U(x,\theta)$ that depends on the normal distance from the boundary $x$ (see Fig. \ref{fig:figure_schematic} -(f)), and on the angle $\theta$ \cite{Bischofs2004}. The equations that govern the dynamics of a cell modeled as an active Brownian particle in an elastic potential are,
\begin{eqnarray}
    \frac{\partial \mathbf{r}}{\partial t} &=& v_{\mathrm{0}} {\mathbf{p}} - \mu_{T}
    \mathbf{\nabla} U + \sqrt{2D_{T}}\bm{\eta}_{\mathrm{T}}(t),  
    \label{eq:Langevin_1_trans}
    \\
    \frac{\partial \theta}{\partial t} &=& - \mu_{R}\frac{\partial U}{\partial \theta} + \sqrt{2D_{R}}\eta_{\mathrm{R}}(t),
    \label{eq:Langevin_1_rot}
\end{eqnarray}
where $D_{R}$ and $D_{T}$ are diffusion coefficients associated with orientational and translational fluctuations of the cell's principal axis and center of mass, respectively, while $\mu_{R}$  and $\mu_{T}$ represent the corresponding rotational and translational mobility. 

For passive bodies in an ambient viscous medium, mobility coefficients depend on the medium's viscosity and temperature, and are coupled via the body's geometry, through the Stokes-Einstein relationship \cite{caraglio2022analytic, di2023active}. Living cells, however, being active and not at equilibrium, do not have to adhere to this constraint. Their self-propulsion velocity can be resolved into a persistent as well as a stochastic part, the latter  arising from random protrusions created by cytoskeletal processes. The persistent self-propulsion $v_{0}$, and random noise in translation, $D_{T}$ and $D_{R}$, are set by subcellular processes in the cytoskeleton, such as the capping, polymerization and depolymerization of actin filaments. In our phenomenlogical model, the details of these processes are ``lumped'' into the few aforementioned motility parameters.
For migratory cells, mobility coefficients arise from dissipative frictional mechanisms at the cell-substrate interface. The friction can contribute additional terms due to memory and inertial effects in the cell dynamics \cite{Bruckner2020}, while the statistics of the cell trajectory may deviate from a persistent random walk \cite{dieterich2008anomalous} in 3D \cite{wu2014three}, effects which we ignore here for simplicity.  We include the effects of stochastic noise via the last terms on the right hand side of Eqs.~(\ref{eq:Langevin_1_trans}) and (\ref{eq:Langevin_1_rot}), $\bm{\eta}_{\mathrm{T}}(t)$ and $\eta_{\mathrm{R}}(t)$ respectively, and correspond to white noise. 
 
The important long-range contribution to the boundary interaction potential $U$, detailed in Appendix App. A, arises from deformations of the elastic substrate.  In addition, we use a short-range steric interaction term to prevent a cell from penetrating the boundary.  The elastic potential arising from the interaction of the cell force dipole, $P_{ij}$ with the substrate deformation, given by the elastic strain $u_{ij}$, it generates in the vicinity of the free or clamped boundary is of the form \cite{Bischofs2003},
\begin{equation}
 U(x, \theta) =-\left(\frac{P^2 }{256\pi E}\right) \frac{f_{\nu}(\theta)}{x^3},
 \label{eq:Potential_full} 
\end{equation}
where $P$ is the strength of the cellular force dipole that is aligned with the cell major axis, parallel to the direction of motility ${\bf p}$, and 
\[
f_{\nu}(\theta)=(a_{\nu}+b_{\nu}\cos ^2 \theta+c_{\nu} \cos ^4 \theta)
\]
%
%
encodes the angular dependence of the potential $U$ that is separable in $x$ and $\theta$ coordinates. We made a simplifying assumption, valid for highly polarized cells such as fibroblasts, by identifying the dipole axis with the direction of motion \cite{yu2017phenomenological, Bose2021}. Substrate elastic properties affect the potential $U$ through its dependence on the Young's modulus $E$, and the Poisson's ratio $\nu$.
Specifically, the angular factor $ f_{\nu}(\theta)$ depends on the substrate Poisson ratio  via constants $a_{\nu}$, $b_{\nu}$ and $c_{\nu}$ (see App. A). Importantly, the constants vary depending on the type of boundary condition - \emph{i.e.}, whether the boundary is free or clamped. Exact forms of these from Ref.~\cite{Bischofs2003} are provided in App. A. 
Subsequently, we use a scaled form of the angular factor defined as 
\[
\Tilde{f}_{\nu}(\theta) \equiv \frac{50}{256\pi}f_{\nu}(\theta),
\] 
such that $\Tilde{f}_{\nu}(\theta) \sim \mathcal{O}(1)$.

The form of the potential in Eq.~\ref{eq:Potential_full} may be rationalized as follows. The force dipole $P_ij$ exerted by the cell interacts with local substrate deformation arising due to the presence of the boundary. This strain field $u_{ij}$ is generated by the associated ``image'' dipole configuration required to satisfy the free or clamped condition on the boundary \cite{walpole1996elastic}. The strain created by a dipole in an elastic half-space decays with distance as $~1/x^{3}$, while it is proportional to the stress given by magnitude dipole moment, $P$. The dipole-dipole interaction potential, $P_{ij} u_{ij}$, therefore scales as $P^{2}/x^{3}$. 
We note that this coarse-grained model of cell traction force distribution as a force dipole is a far-field approximation,valid when the cell-boundary distance is greater than the cell diameter. We further note that the exact form of the potential given by $f_{\nu}$, detailed in App. A, is derived for a dipole embedded in a 3D elastic half-space \cite{walpole1996elastic, Bischofs2004}. We expect a qualitatively similar potential with the same scaling with distance to apply to a cell crawling on the surface of an infinitely thick substrate.

Our model features four dimensionless parameters controlling cell trajectories:
\begin{equation}
\begin{aligned}
    Pe &\equiv \frac{v_{\mathrm{0}}}{D_{R}\sigma},\: A\equiv \frac{1}{50} \left(\frac{\mu_{T}P^{2}} {E D_{R}\sigma^{5}}\right),\:
    B \equiv  \frac{1}{50}\left(\frac{\mu_{R}P^{2}} {E D_{R}\sigma^{3}}\right), \\  D &\equiv 
   {D_{T} \over D_{R}\sigma^{2}}.
\end{aligned}
\label{eq:A_B_Pe}
\end{equation}
The P{\'e}clet number $Pe$ quantifies the relative importance of directed self-propulsion and random motion, and is a measure of persistent motion of the particles in the absence of boundary potential $U$. The  parameter $A$ quantifies the strength of the force, while $B$ quantifies the strength of re-orienting torque, both acting on the cell due to substrate deformation-mediated elastic interactions. Both these elastic interaction parameters depend on the elastic properties of the substrate but are notably independent of active self-propulsion.  
The factor of $1/50$ in the definition of $A$ and $B$ results from the angular average $ \langle f_{\nu}(\theta) \rangle \equiv (1/2\pi)\int_{0}^{2 \pi} f_{\nu} (\theta) d\theta = 1/50$. In this work, we set the substrate Poisson's ratio to a representative  value of $\nu =0.3$ \cite{lo2000cell,li1993new}. $D = D_{T}/D_{R} \sigma^2$ represents the ratio of noise in the translational and orientational degrees of freedom of the cell. Unlike noisy dynamics of thermal origin, these two quantities are not necessarily coupled to each other, and maybe set independently by the cell.

In general, $A$ and $B$ can differ in value depending on the specific mode of cell migration. The ratio $A/B$ is equivalent to ($\mu_{T}/\mu_{R}\sigma^{2}$). For a passive spherical particle at equilibrium in a viscous medium, the ratio $A/B = 1/3$. For elongated rod-like objects, the ratio depends on the aspect ratio and tends to $1/9$ in the limit of infinitesimally thin rods \cite{tao2006isotropic, heyes2019translational, doi1978dynamics}. The case of cells on an elastic substrate is more complex. The values of $A$ and $B$ can strongly depend on the internal mechanisms driving cell motility, an example being internal changes in cell biochemistry that determine the direction of protrusions in the cells. In fact, a high value of $B$ keeps the cell aligned in the direction determined by the elastic potential, and can represent 1D cell migration.

To estimate $A$ in cell culture experiments, we use the typical value for the traction force of a contractile cell adhered to an elastic substrate $F \sim 10 - 100$ \si{nN}, with a distance of $ \sigma \sim 10-50$ \si{\mu m} separating the adhesion sites.
This results in a force dipole moment for a single cell,  $P= F 
\sigma \sim 10^{-12}-10^{-11}$ \si{J} \cite{Bischofs2003}. Using typical values of substrate stiffness in durotaxis experiments, $E \sim 10$ \si{kPa} \cite{duchez2019durotaxis},  rotational diffusion, $D_{R} \sim 10^{-2}$ \si{min^{-1}} \cite{novikova2017persistence}, cell size $\sigma \sim 20$ \si{\mu m}, and previously estimated translational mobility \cite{noerr2023optimal}, $\mu_{T} \sim 0.1 \, \mu\mathrm{m/min} \cdot \mathrm{pN}^{-1}$, we estimate $A \sim 1$.  By changing substrate stiffness and allowing for variation in cell types, we estimate a typical range of $A \sim 0.1-10$, where $A$ can be small on very stiff substrates. Further using $\mu_{R} \sim 10^{-4}$ \si{\mu m^{-1}min^{-1}pN^{-1}}, we estimate $B\sim1$. We again estimate a typical range of $B\sim0.1-10$ by changing the substrate stiffness, where $B$ is small on high substrate stiffness.

We estimate $Pe \sim 0-10$ based on typical cell migration velocities \cite{novikova2017persistence}, $v_{0} \sim 0-100$ \si{\mu m/hr}.
We choose to keep the parameter $D = (D_{T}/D_{R}\sigma^{2})$ fixed at $0$ or $1$ in our simulations. The former simplifying choice corresponds to the regime of highly persistent cell migration characterized by high $Pe$ values, where the effective translational diffusion results from cell reorientations, and is given by $\ v_{0}^{2} /D_{R}$. Finally, we also fix the size of the simulation box to $L=40 \sigma$.

\section{Results}

\subsection{Elastic interactions determine steady state distributions near clamped and free elastic boundaries}
\begin{figure*}[hbt!]
\centering
\includegraphics[width=0.81\linewidth]
{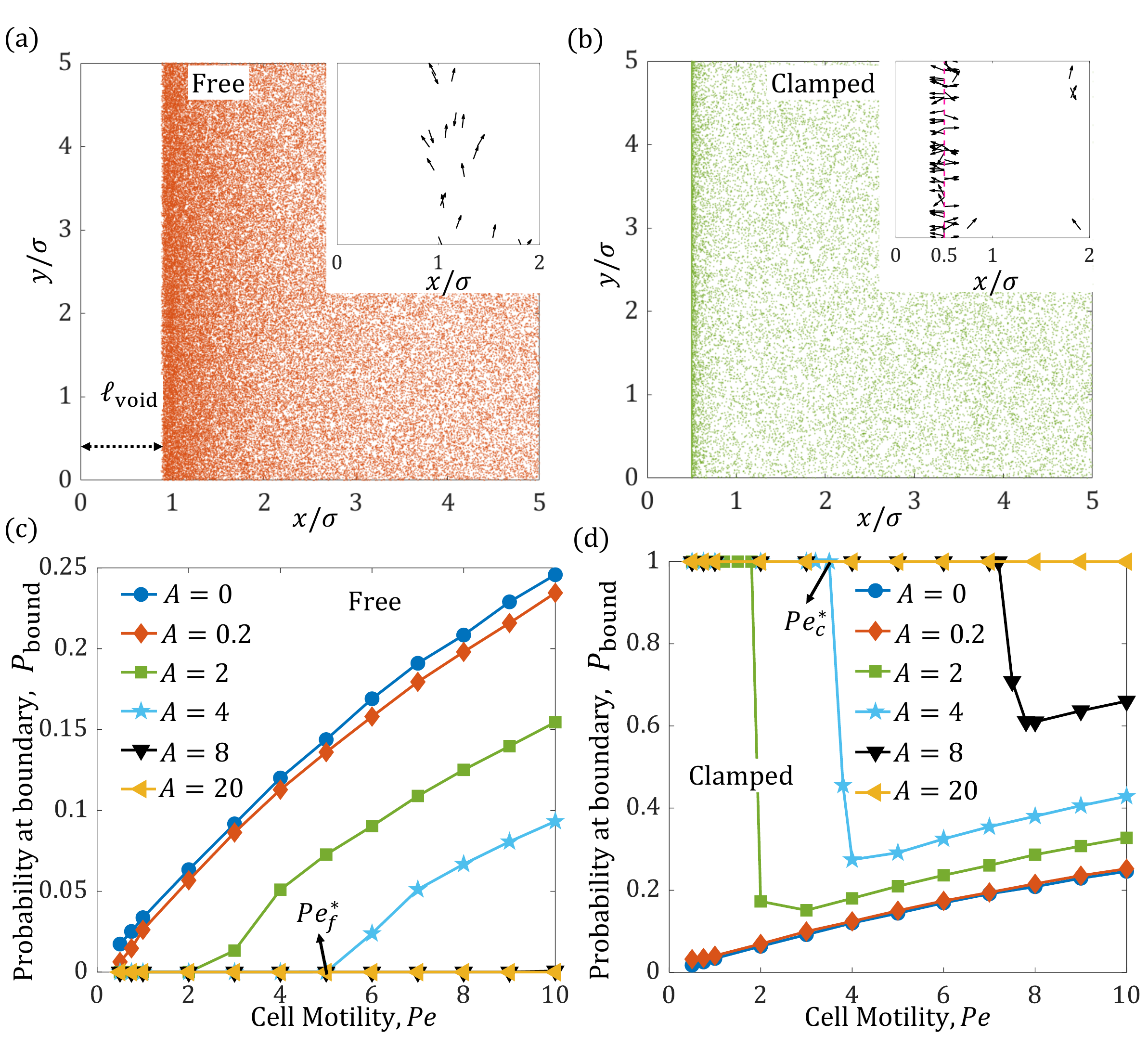}
\caption{ 
\textbf{Spatial probability distribution of single motile cell dipoles near free and clamped elastic confining boundaries.}
Spatial distribution map of model cells for (a) free ($A = B = 16$, $Pe = 8$),  and (b) clamped boundaries ($A = B = 8$, $Pe = 8$), where the data points represent the occurrence of cells at  corresponding positions, sampled at regular intervals from multiple simulation trajectories. Insets show a magnified view close to the boundary, at $x_{b} = \sigma/2$, with arrows indicating the orientation of the self-propulsion of the cell at each sampled position in its trajectory. 
In (a), the repulsive potential from the free boundary results in a void region of extent $\ell_{\mathrm{void}}$, which cells are unable to access. Cells close to the boundary are oriented parallel to it due to elastic torques (see inset). In (b), the attractive potential from the clamped boundary causes accumulation of cells   
while the elastic torque orients the cells perpendicular to the boundary (inset in (b)). In the inset in (b), the pink dashed line shows the center of all the cells at $x = 0.5\sigma$.
(c,d) 
We quantify the accumulation as the measured probability of cells, $P_{\mathrm{bound}}$, being at the boundary for free (c) and clamped (d) conditions, for various values of $Pe$ and $A (=B)$. (c) The localization near a free boundary decreases with increasing elastic repulsion $A$, 
but increases with $Pe$ due to motility-induced accumulation. 
$Pe^*_f$ corresponds to the threshold value of motility at which the cell's self-propulsive force can overcome the repulsive boundary force and reach the boundary for given value of $A$ (marked here for $A = 4$). For $Pe < Pe^*_f$, there are no cells at the boundary, \emph{i.e.} $P_{\mathrm{bound}}=0$, and a void region develops.  
(d) The localization at a clamped boundary increases with $A$, while there are two regimes of dependence on $Pe$. At low $A$, $P_{\mathrm{bound}}$ increases with $Pe$, since faster cells reach the boundaries more frequently and spend more time there. At high $A$ and low $Pe$, cells are trapped at the boundary by the attractive elastic force leading to $P_{\mathrm{bound}} = 1$. There is a sharp drop in $P_{\mathrm{bound}}$ at a threshold value $Pe^*_c$, at which self-propulsion can overcome the elastic attraction. 
For $Pe > Pe^{*}_{c}$,  $P_{\mathrm{bound}}$ increases with $Pe$ due to motility-induced accumulation. 
}
\label{fig:figure_boundary}
\end{figure*}
Cells migrating through their complex extracellular matrix sense and respond to physical cues \cite{charras2014physical}. They are expected to respond to both gradients in substrate stiffness and confining boundaries.
Theoretical models describing the statistical behavior of active particles under confinement have been studied extensively in earlier works. These works compute the density, surface density, polarization, and orientation distributions of active particles between two parallel confining boundaries or at straight or curved boundaries \cite{ezhilan2015distribution, yan2015force, elgeti2015run, wensink2013differently, wagner2017steady}. These studies show that statistical steady state distributions depend strongly on  particle activity, the shape of the particles, and the curvature of the boundaries. Passive particles moving in a constant temperature, non-deforming medium without persistent self-propulsion ($Pe = 0$), are expected to reach thermodynamic equilibrium and have uniform distribution between the boundaries that maximizes entropy.  In contrast, as $Pe \to \infty$, particles populate the boundaries at all times with the probability of finding particles at the boundary tending to unity resulting in a diverging surface density. The surface density also depends on the curvature of the surface \cite{fily2017mechanical}.

Cell-boundary interactions mediated by an ambient material medium have also been investigated in detail for a related class of microswimmer problems, including the interaction of low Reynolds number microswimmers such as bacteria, algae and sperm with boundaries \cite{Drescher2011, Lopez2014, tailleur2009sedimentation, berke2008hydrodynamic, smith2009human, giacche2010hydrodynamic, li2011accumulation, berg1990chemotaxis, lauga2006swimming, lemelle2010counterclockwise, patteson2016active}. Unlike the animal cells studied here that act as contractile dipoles, free swimming organisms can act as pushers (bacteria, and sperm) or pullers (algal cells). Far from interfaces, pushers generate extensile force dipoles on the ambient fluid, while pullers exert contractile force dipoles. Additional stresses on the fluid are generated in pushers due to ``rotlet'' dipoles arising from counter-rotation of the cell body and the flagellar bundle. The presence of interfaces near swimming cells results in wall induced forces and torques on these swimmers; these effects arise due to the requirement that the overall fluid fields generated by the moving cells, and mediated by the interface(s), satisfy appropriate boundary conditions -- that is no-slip for solid walls, or stress-free for free surfaces. 
Experimental studies on swimmers near surfactant-free, solid, no-slip surfaces indicate that, irrespective of the type of dipolar swimmer, microorganisms tend to accumulate near the interface albeit with varying orientations. Pushers tend to align parallel to no-slip solid interfaces due to hydrodynamic torques, and swim along the surface exhibiting long residence times \cite{Drescher2011,lauga2006swimming}. Analyzing the competition between cell-wall hydrodynamic attraction and rotational diffusion, Drescher {\it et al.} estimated characteristic cell-wall interaction time scales and deduced that hydrodynamic wall-induced attraction dominates provided the distance from the wall $x < P(a/v_{0}D_{R})$ where $\sigma$ is the cell (body) size, $P$ is the hydrodynamic dipole strength, and $v_{0}$ is the self propulsion speed. Contractile pullers meanwhile have been observed to align perpendicular to the interface and remain trapped until they can reorient and escape due to thermal noise or rotational diffusion arising from variations in the swimming mechanism \cite{Drescher2011, Schaar2015}. 
Interestingly, pushers are theorized to be attracted to surfactant-free (clean) interfaces with the Stokes dipole oriented and aligned parallel to the interface, for both free surfaces as well as for solid walls \cite{lauga2006swimming, li2011accumulation}. 

In this work, we investigate the effects of cell-interface elastic and steric interactions on the boundary and bulk distributions of active particles representing motile cells on elastic substrates. Motivated by the process of single cell durotaxis across sharp gradients of substrate stiffness as shown in Fig.~\ref{fig:figure_schematic}a , we study the effect of elastic forces and torques on the density and orientational distributions of motile cells at the confining boundary. We carry out simulations of cell trajectories using the model Eqs.~\ref{eq:Langevin_1_trans}-\ref{eq:Langevin_1_rot} for a range of values of self-propulsion, $Pe = 0.5 - 10$, and elastic interaction strength, $A (= B) = 0$ to $20$, that were estimated in the  model section for cell culture experiments. 
From these simulations, we compute the probability of finding a particle at the boundary using 
$P_{\mathrm{bound}} = N_{\mathrm{bound}}/{N_{\mathrm{total}}}$, where $N_{\mathrm{bound}}$ is the number of occurrences of the particle at the boundary -- that is, its center is located at $x= x_{b} \equiv \sigma/2$ {\em after} the instantaneous displacement/reassignment step (Supp. Note 3). ${N_{\mathrm{total}}}$ meanwhile is the total number of times the particle is observed. 

To aid in the analysis and interpretation of results, we set $D=0$, that is we switch off translational diffusivity $D_{T} = 0$, in our simulations. In the short time limit relative to the persistence time $D_{R}^{-1}$, this allows cells to localize and stay at the boundary except when the directed self-propulsion drives them away. Over longer times however, an effective diffusivity that is $v_{0}^{2}/D_{R}$ arises due to the combination of self-propulsion and re-orientations represented by rotational diffusion. 

As a point of departure, we first describe the results in the absence of elastic interactions with the boundary, $A = B = 0$. Geometric confinement prevents cells from leaving the system in the direction normal to the boundaries. Consistent with previous studies on non-interacting Active Brownian Particles (ABPs) \cite{ezhilan2015distribution}, we observe localization of cells at the boundaries, with the associated number densities at the boundaries ($N_{\mathrm{bound}}$) increasing with the P{\'e}clet number ($Pe$). To rationalize this, we note that increasing $Pe$ is equivalent to faster cell migration speed and more persistent motion (Fig. \ref{fig:figure_boundary} (c), (d)). Cells are able to translate over longer distances due to decreased effects of diffusion. Once the cells reach the boundaries however, they tend to remain there since they are oriented towards the boundary, until reorientation is caused by rotational diffusion over the characteristic timescale $\sim D_{R}^{-1}$. Upon reorientation, the cell's orientation given by the polarization vector's angle is pointed away from the boundary, $\theta < \pi/2$. If the cell's self-propulsive force is strong enough to overcome the elastic attractive force, the cell escapes from boundary trapping and moves back into the bulk.  
Increasing cell $Pe$ decreases the time spent between the confining boundaries which in turn increases their probability to be at the boundary. 

Such localization at the boundary, while well-known for microswimmers (as previously described), and also for synthetic active particles, is yet to be demonstrated for crawling animal cells. We propose that this effect may be detected by tracking spatial probability of cells in a dilute cell culture experiment where confinement is created by micro-patterning the underlying elastic substrate into two discrete regions, only one of which favors adhesion. The interface between these two regions will act as a confining boundary that restricts cell migration into the unfavorable region where cells cannot adhere. Henceforth in this work, we term this increased localization of cells at the confining boundary by purely kinetic means, motility-induced accumulation (MIA). 

The probability of a cell being at the boundary is strongly modulated by the nature of elastic interactions in our model. Specifically, the sign of elastic interaction depends on the type of boundary condition, clamped (i.e. ``no displacement'') or free (i.e., ``no stress''). For stress-free boundary conditions representing an interface with a softer substrate, increasing repulsive forces act on the cells as they approach the boundary. Therefore in this case, cells are unable to reach the boundary and remain a distance away from it, see Fig.~\ref{fig:figure_boundary}(a). Furthermore, the torque from the elastic interaction induces cells close to the boundary to align parallel to it, see inset to Fig.~\ref{fig:figure_boundary}. Increasing the interaction parameter $A$ (here we set $B=A$) increases the length of the region over which the repulsive force acts and reduces the probability of a cell being at the boundary.  For $A > 0$ and low $Pe$, there is no localization at the boundary, Fig.~\ref{fig:figure_boundary}(c). Quantifying this localization by a probability density of observing particles at the boundary 
we find from our simulations that for each value of $A$, there exists a critical P{\'e}clet number $Pe_{\mathrm{f}}^*$ at which the localization probability, $P_{\textrm{bound}}$ at the boundary becomes non-zero.  For $A > 0$, increasing the P{\'e}clet number to values larger than $Pe_{\mathrm{f}}^*$, increases the probability of the cells to localize at the boundary. When  $Pe < Pe_{\mathrm{f}}^*$, cells cannot reach the boundary resulting in a void region evident in Fig.~\ref{fig:figure_boundary} (a). We find that $Pe_{\mathrm{f}}^*$ increases with the interaction parameter $A$. This increase is expected to be linear from force balance. 

The situation is quite different for cells interacting with clamped boundaries. In this case, cell-boundary elastic interactions are attractive and increasing $A$ localizes more cells at the boundary,  Fig.~\ref{fig:figure_boundary}(b). In addition, the elastic torque due to the boundary orients cells perpendicular to the boundary, seen in Fig.~\ref{fig:figure_boundary}(b) (inset). At low values of $A$ (for $A < 2$), we find that $P_{\mathrm{bound}}$ increases monotonically with $Pe$. This is a consequence of the enhanced flux towards the boundary due to the higher speed ($Pe$), and the attractive potential that traps the cells. For higher $A$ ($A > 2$), and at low $Pe$, cells  are strongly localized at the boundary with $P_{\mathrm{bound}} = 1$ due to the strongly attractive elastic force from the clamped  boundary. For $Pe \geq 1$, we see a reduction in $P_{\mathrm{bound}}$ as escape from the boundary is increasingly facilitated by the greater speed. The critical P{\'e}clet number $Pe_{\mathrm{c}}^*$ at which the cells overcome the attractive interaction with the clamped boundary and move into the bulk increases with $A$ and is expected to be linear from force balance. Eventually however as $Pe \gg 1 $, the role of the elastic potential becomes subdominant to the effects of increased motility, and particles are more likely to be observed at the boundary than in the bulk. 
\begin{figure*}[ht]
\centering
    \includegraphics[width=0.81\linewidth]{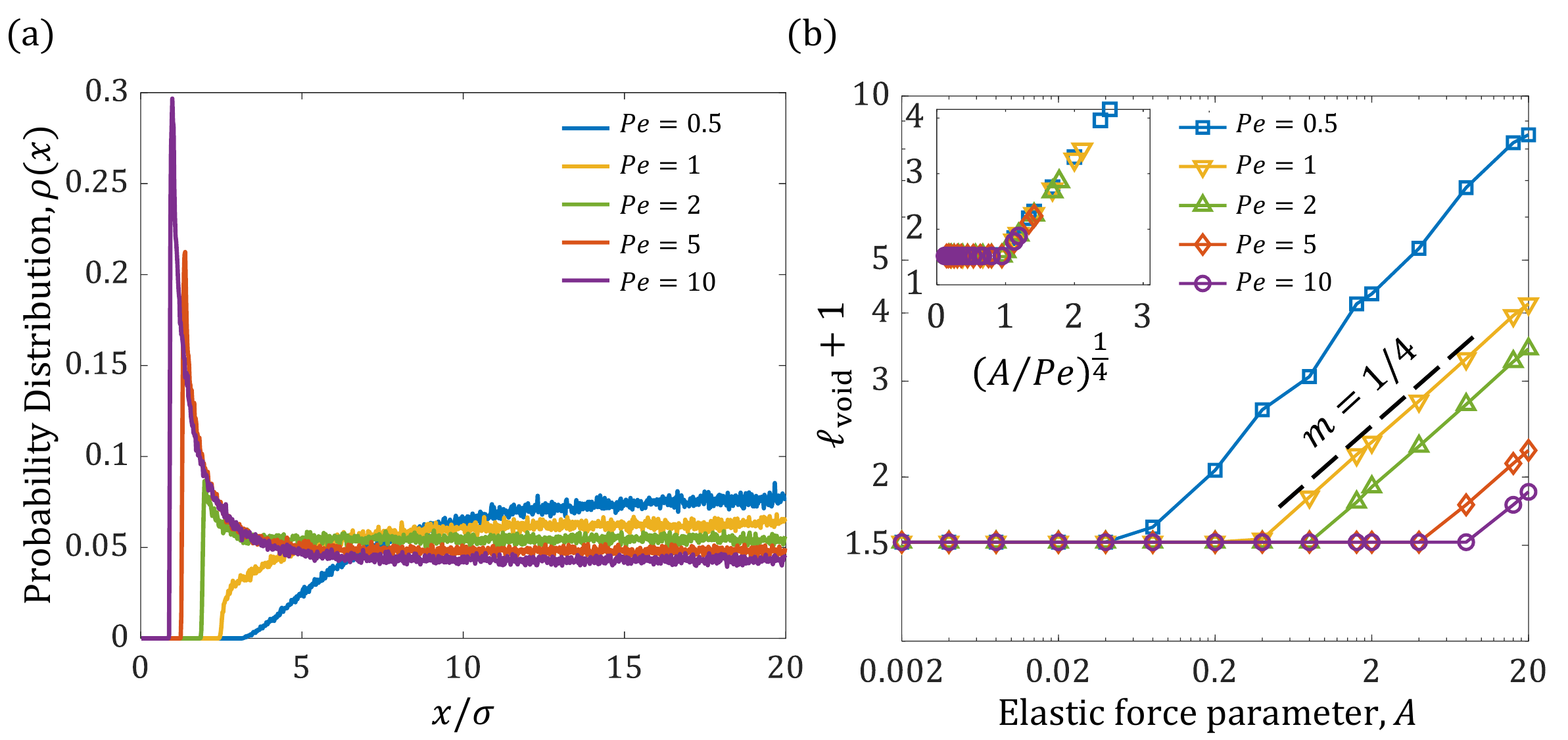}
    \caption{
    \textbf{Void region near a free boundary, and its dependence on motility and elastic interactions.}
    (a) The spatial probability density profile $\rho(x)$ of cells as a function of distance $x$ from the free boundary is shown for $A = 20$ with $Pe = 0.5, 1, 2, 5, 10$. Increasing $Pe$ leads to decreasing void region. (b) The void length scales as $ \sim A^{1/4}$, and $ \sim Pe^{-{1/4}}$ (for constant $A$) as predicted from force balance,see Eq.~(\ref{eq:void}). Inset shows the collapse of the $\ell_{\mathrm{void}} + 1$ vs $\left(A/Pe\right)^{1/4}$. 
    }
    \label{fig:figure_void}
\end{figure*}
In contrast, for a clamped elastic boundary, when the strength of  the elastic attraction $A$ is sufficiently larger than the persistent cell motility $Pe$,  $P_{\mathrm{bound}} = 1$ implying cells are strongly localized at the boundary. These cells have a higher chance of crossing over to the stiffer side. On the other hand, an elastic free boundary decreases $P_{\mathrm{bound}}$ thereby reducing the cells' tendency to go towards the softer substrate. Both these types of interactions from clamped and free boundaries, while distinct, promote durotaxis. On the other hand, higher cell migration speeds promote their motility-induced accumulation at a confining boundary without discriminating between stiffer and softer substrates.

\subsection*{Free elastic (repulsive) boundary induces depletion and prevents anti-durotaxis} 
We have demonstrated that our simulated cells are repelled by the free boundary due to the nature of the elastic potential. We track the positions of all cells over time and establish the closest distance from the boundary accessed by each. We showed in Sec. IIA that the repulsive force from the free boundary induces a effective void region where cell do not penetrate, see Fig.~\ref{fig:figure_boundary}(a). 

To characterize this void region systematically, we plot the statistically attained (time averaged and ensemble averaged for all cells ) probability distribution function $\rho(x)$ as a function of $x$ (the distance from the boundary) for various values of $A$ and $Pe$. To obtain $\rho(x)$, we simply record the positions of the cells after sufficient time required to reach steady state has elapsed.  The length of the void region $\ell_{\mathrm{void}}$ is evaluated through these distributions, and is measured as the minimum distance at which the spatial density attains a non-zero value. 
For fixed values of A (for instance, $A$=20 in Fig.~\ref{fig:figure_void} - (a)), we find that increasing $Pe$ decreases the length of the void region. In general, increasing $A$ increases $\ell_{\mathrm{void}}$ while increasing $Pe$ decreases it.

We estimate $\ell_{\mathrm{void}}$ for $Pe$ from $0.1$ to $10$ and for $A=B$ from $0.002$ to $20$ to discern  trends from physical scaling. Consider the balance of forces acting on a cell located at $x = \ell_{\mathrm{void}}$. 
Balancing the self-propulsion ($\sim Pe$) and elastic interaction forces ($\sim A/(x/\sigma+1)^{4}$) that move the cell, we obtain 
\begin{equation}
(\ell_{\mathrm{void}}/\sigma) + 1 \sim \left({A/Pe}\right)^{\frac{1}{4}}. 
\label{eq:void}
\end{equation}
Indeed, void lengths extracted from simulated probability distributions  confirm this theoretically predicted scaling in Fig.~\ref{fig:figure_void}(b).  Experimentally, the presence of a void region may be detected by culturing and tracking cells on a  stiff adhesive region of an elastic substrate, adjoining a very soft, non-adhesive region that acts as a free boundary. Our model predicts low probability of finding cells in a void region.

\subsection*{Clamped (attractive) boundary induces durotactic trapping}
In our model, the clamped boundary condition represents the cell being on the softer substrate. This configuration facilitates durotaxis by inducing an attractive force and aligning torque on the cellular force dipole. Such cells therefore tend to be trapped at the confining boundary. Since cell migration is stochastic and not deterministic, they can sometimes go opposite to the durotactic direction. This is possible in our model through reorientation via rotational diffusion, which represents random internal fluctuations in cell polarity. Once the cell reorients and points away from the confining boundary, it can escape from the trapped state if the self-propulsion is strong enough to overcome the elastic attraction, $Pe \gtrsim A$. 

\begin{figure*}[ht]
\centering
\includegraphics[width=0.81\linewidth]{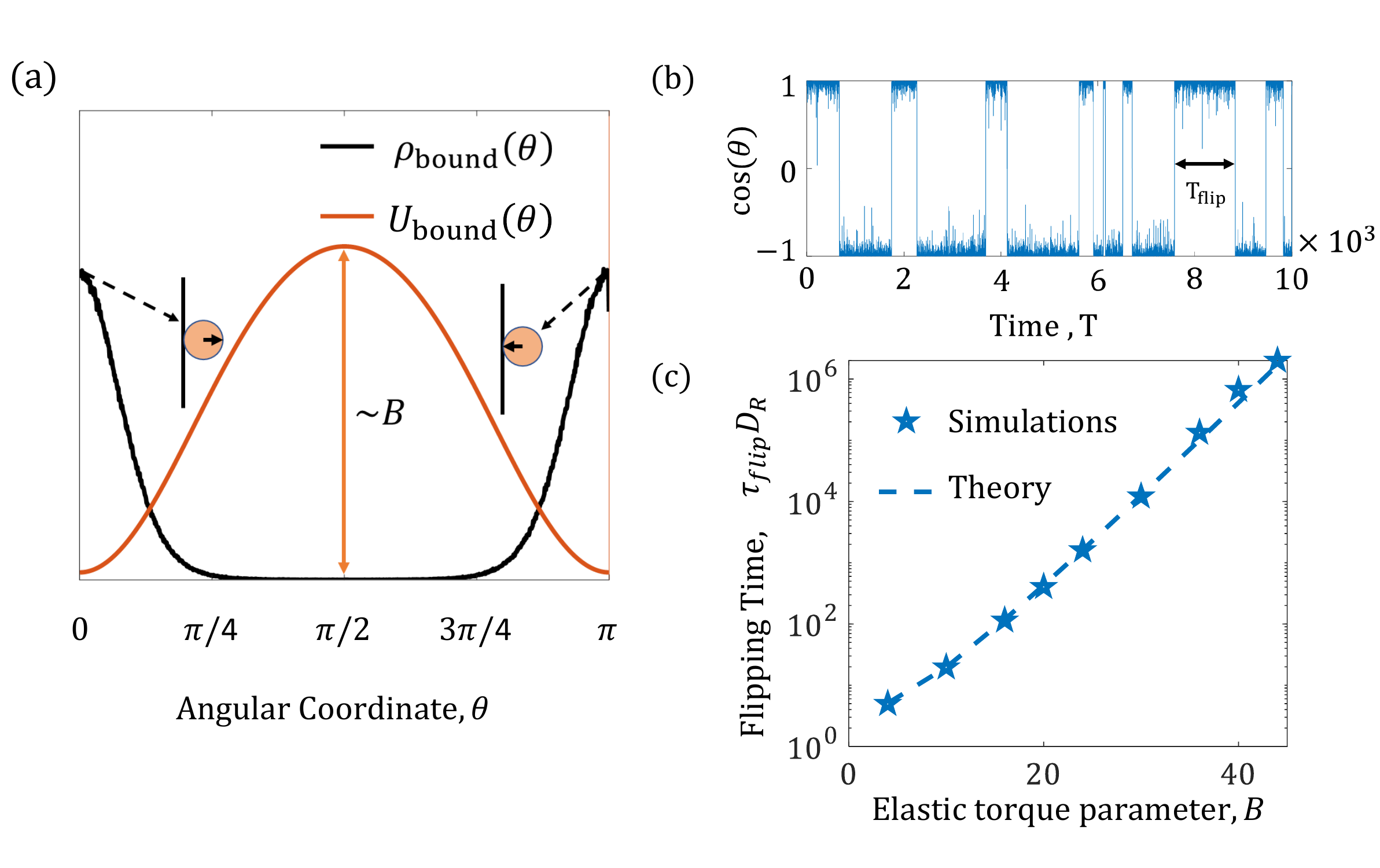}
\caption{\textbf{Cell reorientation (flip) kinetics  at clamped (attractive) boundary predicted by barrier crossing theory.} (a) For a clamped boundary, and at very high values of $A$ and $B$, cells localize at the boundary 
and experience a boundary elastic potential with minima at $\theta=0$ and $\theta = \pi$, corresponding to pointing away from or towards the boundary, respectively.
(b) Rotational diffusion enables the cell to transition between these orientation states. These random flips are recorded for a cell stuck at the boundary for $A = B = 20$. (c) The average time interval between these flips is measured as a function of the elastic torque parameter, $B$. The flipping time follows Kramer's theory of barrier crossing given by Eq.~(\ref{eq:flip_time}).
}
\label{fig:figure_flip}
\end{figure*}
When $A \geq 1$ and $Pe \sim 1$, cells tend to localize at the clamped boundary, as seen in Figs.~\ref{fig:figure_boundary}(b) and (d). At the same time, a large elastic torque, $B \geq 1$, orients the direction of propulsion directly towards or away from the boundary, as shown in the schematic Fig.~\ref{fig:figure_flip}(a). 
We now quantitatively investigate the rate at which the cells trapped at the boundary flip their orientation from pointing towards the boundary to pointing away from it, or \textit{vice versa}. This provides an estimate of the time scale over which cells can remain trapped at the boundary. Since reorientation dynamics is dominated by the boundary-induced elastic torque, we focus on $B$ as our parameter of interest in this subsection. Since escape after rotation diffusion-enabled reorientation is possible through persistent motility alone when $Pe > A$, we continue to keep the translational diffusion parameter $D=0$ in this section.

 \begin{figure*}[ht]
    \centering
    \includegraphics[width=0.93\linewidth]{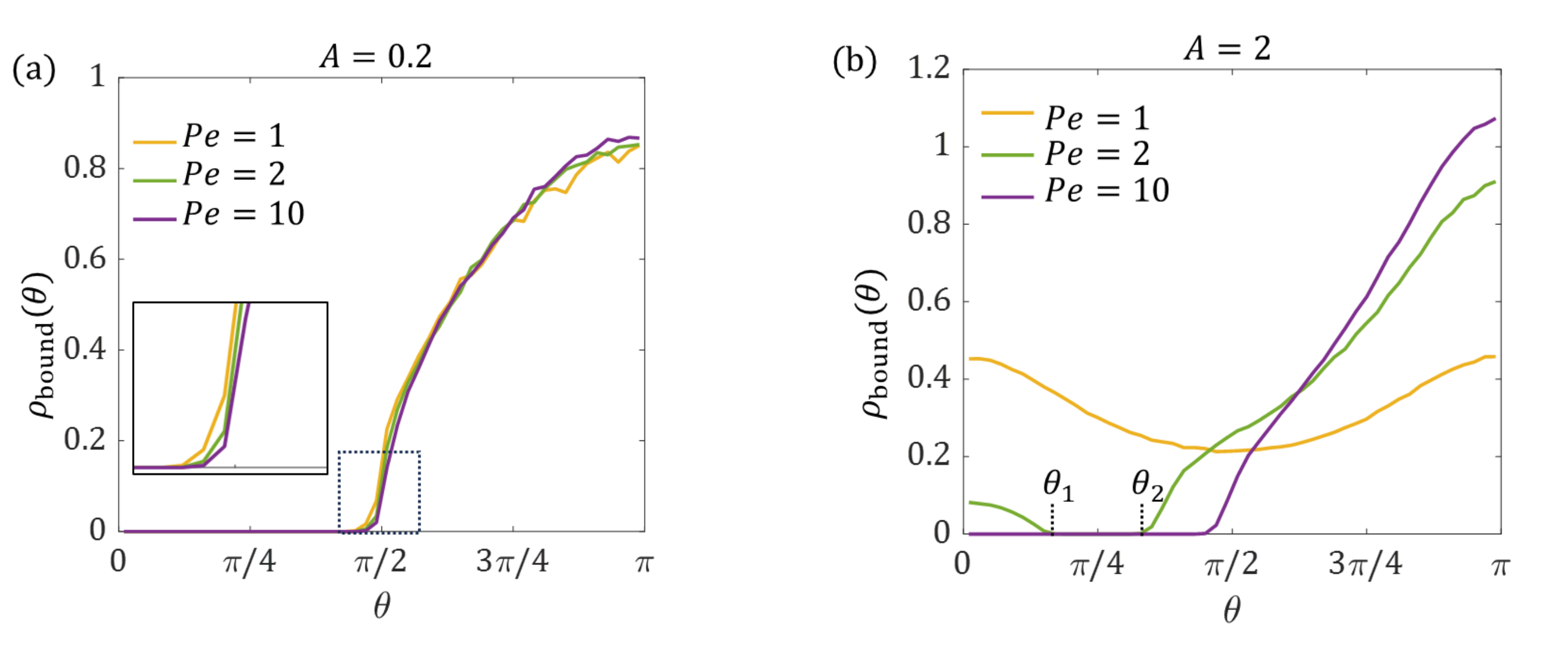}
    \caption{ \textbf{Orientational probability density of cells at clamped boundary indicates extent of trapping.} 
    (a) For $B = A = 0.2$, cells are weakly attracted and oriented by the boundary. The cell can then escape even when it has a small component of motility pointing away from the boundary, resulting in $\rho_{\mathrm{bound}}(\theta)=0$ when $\theta < \theta_{\textrm{esc}}$. Increasing $Pe$ widens the cone of escape (inset (a)), with  $\theta_{\textrm{esc}}$ being $77.4^{0}$, $81^{0}$ and $84.6^{0}$ for $Pe = 1$, $2$ and $10$ respectively. (b) When elastic attractive force is larger, $A = B = 2$, we identify 3 distinct regimes that are $Pe$-dependent. For low $Pe = 1$, the cells are trapped at the boundary but free to reorient due to rotational diffusion, remaining preferentially perpendicular to the boundary. At high $Pe = 10$, the cell motility can overcome the boundary attraction when the cell is oriented away at some $\theta_{\textrm{esc}}$, like (a). At intermediate $Pe = 2$, the cells are only able to escape when their orientation lies in a range between  $\theta_{1} = 30.7^{0}$ and $\theta_{2} = 55.8^{0}$.}
    \label{fig:figure_ori}
\end{figure*} 

 In the limit of large elastic torque parameter $B \gg 1$, cells at the boundary are always oriented perpendicular to the boundary, pointing towards or away from it. As depicted in Fig.~\ref{fig:figure_flip}(a), a cell can thus reside in one of two possible orientation states: {either pointing towards the boundary ($\theta = 0$), or away from ($\theta = \pi$) the boundary}. These two states are  the minima of the potential double well in orientation, $U(x=x_{b},\theta)$. Flips are defined as the large, stochastic, reorientation events caused by rotational diffusion when $\theta$ changes from $\pi$ to $0$ or \textit{vice versa}. To measure the average frequency of flips, we track the change in orientation of cells localized at the boundary, given by the angle $\theta$, see Fig.~\ref{fig:figure_schematic}(c). Thus, flips result in change in sign of $\cos{\theta}$ , seen in  Fig.~\ref{fig:figure_flip}(a). A typical simulation trajectory in Fig.~\ref{fig:figure_flip}(b) shows that flipping occurs multiple times during a given simulation run, even at high values of $B$.
 We define and measure a time taken by a cell to flip, $\tau_{\mathrm{flip}}$, as the residence time of the cell in either state. Following the orientation of a single cell over the time it is trapped at the boundary provides a distribution of flipping times. In Fig.~\ref{fig:figure_flip}(c), we show the mean flipping time $\tau_{\mathrm{flip}}$, averaged over many cell trajectories, for a range of large $B$ values ($B=4-45$) with $A$ set to equal $B$. The dependence of $\tau_{\mathrm{flip}}$ on $B$ follows the predicted form of  Kramer's theory of barrier crossing \cite{Kramers1940}, 
\begin{equation*}
    \tau_{\mathrm{flip}} = \frac{2\pi}{\mu_{R}\sqrt{U''(0)|U''({\pi/ 2})|}} \exp{\left(\mu_{R}\frac{U({\pi / 2}) - U(0)} {D_{R}}\right)} 
  \end{equation*}
    \begin{equation}
  \sim \frac{1}{B} \exp{\left(B\frac{(b_{\nu}^{c} + c_{\nu}^{c})}{864\pi}\right)}.
\label{eq:flip_time}
\end{equation}
In deriving this equation, we used the form of the elastic potential $U(x, \theta)$ given in Eq.~\ref{eq:Potential_full}. Note that since this simulation is for cells trapped at the boundary that are free to change orientation, the potential $U(x,\theta)$ is evaluated at a fixed value of $x=\sigma/2$.  The theoretically predicted flipping times from Eq.~\ref{eq:flip_time} (dashed line) closely agree with the simulation data in Fig.~\ref{fig:figure_flip}(c).

For low or moderate values of $B$ however, cells at the boundary may adopt orientations other than just $0$ and $\pi$.  This is captured by  the steady state orientational probability distribution $\rho_{\mathrm{bound}}(\theta)$ of the cells at the boundary, shown in Fig.~\ref{fig:figure_ori} for two representative values of $B$.  At $A =B = 0.2$, Fig.~\ref{fig:figure_ori}(a), both force and torque from the elastic interactions with the boundary are low. Cells pointing away from the boundary with $\cos{\theta > 0}$ are not strongly attracted by the boundary and may move away self-propulsion. The angle at which these cells lose contact with the boundary, defined here as $\theta_{\textrm{esc}}$, is then the minimum angle  at which $\rho_{\textrm{bound}} (\theta)$ just becomes non-zero. There is no probability of finding cells at the boundary with orientation, $\theta < \theta_{\textrm{esc}}$ at steady state, because these cells have escaped back into the bulk. 
In this small $B$ regime, the escape angle is close to, but smaller than $\pi/2$. Increasing $Pe$ increases the $\theta_{\textrm{esc}}$ slightly towards $\pi/2$, as shown in the inset to Fig.~\ref{fig:figure_ori}(a).\\

For moderate values of $B$, such as when $B =A = 2$, we observe three distinct regimes separated by two transition P{\'e}clet numbers, $Pe_{1}$ and $Pe_{2}$, as seen in Fig.~\ref{fig:figure_ori}(b).
All results in Fig.~\ref{fig:figure_ori}, including the three possible behaviors in  Fig.~\ref{fig:figure_ori}(b),  may be quantitatively understood from a simple force balance argument. In these simulations without translational diffusion ($D=0$), a cell can escape from the boundary only if the attractive force from the boundary is overcome by the normal component of its self-propulsive force. 
Evaluated at the boundary position, $x=x_{b}=\sigma/2$, this force balance has the form 
\begin{equation}
    Pe \cos \theta    = 
    \frac{3A}{\left(x_{b}/\sigma + 1\right)^4}\Tilde{f}_{\nu}(\theta)
\label{eq:force_bal_boundary}
\end{equation}
where $\Tilde{f}_{\nu}(\theta)$ is the rescaled form of $f_{\nu}(\theta)$ in Eq.~\ref{eq:Potential_full}, such that $\Tilde{f}_{\nu}(\theta) \sim 1$. 
The conditions for the existence of solutions of this force balance equation (Supp. Note 3) determine three possible regimes of the orientational distribution of trapped cells.  For low values of  $Pe \leq Pe_{1}$, the elastic attractive force from the boundary, given by $A$, is strong enough to keep cells trapped at the boundary, even when the cell is oriented away from it. 
At high P{\'e}clet number, $Pe > Pe_{2}$, cells are able to overcome  the boundary attraction provided the orientation angle $\theta < \theta_{\textrm{esc}}$, where $0 < \theta_{\textrm{esc}} < \pi/2$.  At intermediate P{\'e}clet numbers, $Pe_{1} < Pe < Pe_{2}$,there exists a range of orientation angles, $0< \theta_{1}$ to $ \theta_{2} < \pi/2$,between which cells can escape. If $\theta < \theta_1$, the attractive force from the boundary is too strong and if $\theta > \theta_{2}$, the cell leans towards the boundary and cannot propel away. Thus, there is an angular cone of escape between $\theta_1$ and $\theta_2$. 

When $A=2$, we estimate $Pe_{1}=1.82$ and $Pe_{2}=2.14$, respectively.  This corresponds to the results in Fig.~\ref{fig:figure_ori}b, for $B=A=2$, where all the three regimes discussed above occur. The cells with $Pe =1 < Pe_{1}$ are trapped at the boundary for all orientations. The orientational distribution has peaks at $\theta =0$ and$\theta = \pi$, corresponding to the minima of the boundary potential, $U_{\textrm{bound}}(\theta)$. Those with the intermediate $Pe_{1}< Pe =2 < Pe_{2}$ exhibit a finite range of orientations where the probability density vanishes. Cells with high $Pe = 10 > Pe_{2}$ can escape at all angles higher than a $\theta_{\textrm{esc}}$ near $\pi/2$.  This last case is observed at all $Pe$ values shown for $A=0.2$ in  Fig.~\ref{fig:figure_ori}a, since the theoretically estimated values of $Pe_{1}$ and $Pe_{2}$ from the analysis in Supp. Note 4 are $0.182$ and $0.214$ respectively. Thus, the force balance in Eq.~\ref{eq:force_bal_boundary} and resulting self-propulsion-dependent escape criteria quantitatively explain our simulated orientational distributions for cells trapped at the boundary in Figs.~\ref{fig:figure_ori}. 
 
If the elastic force from the boundary is very strong, i.e., $A \gg Pe$ the cells cannot escape the influence of the boundary and will all participate in durotaxis.  Escape is likelier when the gradient in substrate stiffness is small, such that the boundary attractive force and the cell's active propulsive force are comparable. The rotational diffusion in our model corresponds to random protrusions and internal chemical signaling that can reverse the polarization of the cells,  while the propulsion drives them away from the boundary. 

{Three mechanisms influence the motion of cells - elastic interaction forces, self-propulsion, and random motion. For zero to very small $Pe$ numbers, we expect random motion to dominate over the deterministic self-propulsion force.  Balancing elastic interaction energy in the vicinity of the clamped boundary with effective thermal energy gives us 
$P^2/(E \ell_{\mathrm{E}}^3) \sim D_{R} \sigma^{2}/ \mu_{R}$, or $\ell_{\mathrm{E}} \sim \sigma A^{1/3}$, the length-scale quantifying the distance from the boundary for which elastic interactions dominate. For $A$ in the range $1-10$, we find that $ \ell_{\mathrm{E}}/\sigma $ varies from $1$ to $\approx 2.15$. 
For moderate to large P{\'e}clet numbers, the relevant balance now comes from the competition between the attractive elastic force, and the self-propulsion force. In this case, we find $\ell_{\mathrm{E}} \sim \sigma (A/Pe)^{1/4}$. We note that the propulsion force may not always act in parallel to the elastic force. Nonetheless, when $x < \ell_{\mathrm{E}}$ elastic forces win and the net force moves cells towards the boundary. When the typical cell spacing is larger than these elastic interaction length scales, as expected for dilute cell cultures, our single cell model will apply.}

Our predictions for the orientational distribution and dependence of reorientation (flipping) timescales in Fig. ~\ref{fig:figure_ori} may be checked in experiment by tracking the orientation and polarization (\emph{i.e.} the direction of migration) of cells cultured on elastic substrates. How these quantities depend on on $A$ and $Pe$ may be checked by performing experiments on substrates of varying stiffness and quantifying cell traction (related to $A$) and migration speed (related to $Pe$).

\subsection*{Comparison with experiment and predicted durotactic phase diagram}
\begin{figure*}[ht]
\centering 
\includegraphics[width=0.81\linewidth] {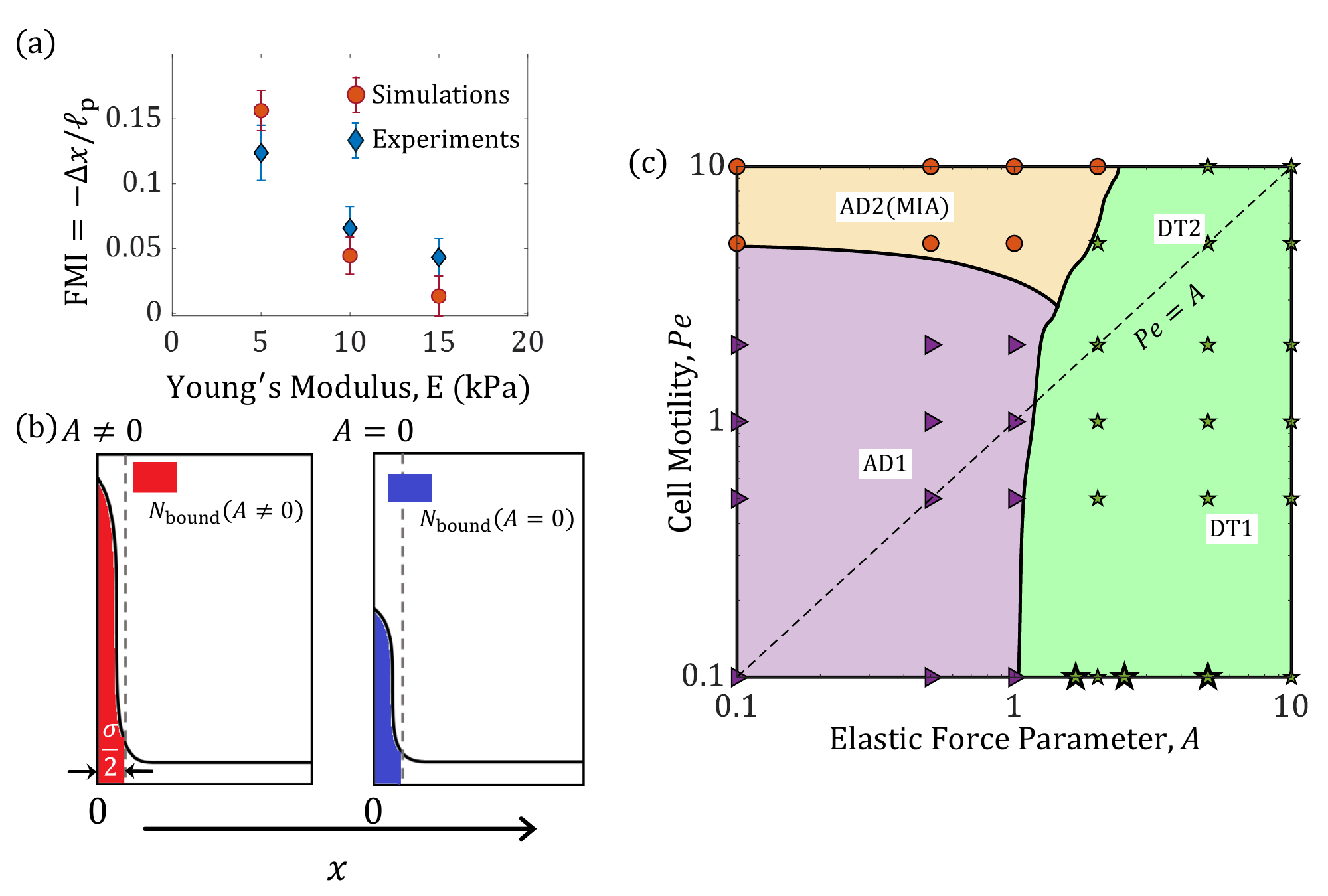}
\caption{{\bf Comparison of cell migration index with experiment and a predicted durotactic phase diagram.} 
(a) The forward migration index defined as the ratio of cell displacement towards the boundary and its total path length, FMI  $= -\Delta x/\ell_{p}$, is calculated from simulations at $Pe = 0.1$ and $D = 1$. 
Simulation results (blue diamonds) compare well with experimental data (orange circles) obtained by DuChez et al. for U-87 gliblastoma cells on an elastic substrate with gradient in stiffness from $2-18$ kPa \cite{duchez2019durotaxis}. The substrate had three different stiffness regions with effective Young's modulus of $5$ kPa (soft), $10$ kPa (medium) and $15$ kPa (stiff).
(b) To calculate the value of durotactic index (DI, defined in Eq.~(\ref{eq:DI})), we simulate and compare the number of cells trapped at a confining boundary for $A \neq 0$ with the corresponding $A=0$ case at the same $Pe$ value. 
(c) Simulated phase diagram in $A-Pe$ space classified according to durotactic index and boundary accumulation. The durotactic region (green) corresponds to simulated cells (green pentagrams) with a DI greater than a threshold value (DI$> 0.27$), which corresponds to the DI value of cells at $A = 1$, escaping through random diffusive motion. The $Pe = A$ line separates the durotactic region into the diffusion-dominated regime (DT1) and motility-dominated regime (DT2). The cells with DI smaller than the $A=1, Pe=0$ case (DI$<0.27$) are classified as adurotactic, AD1 (purple), or adurotactic with motility-induced accumulation, AD2(MIA) (orange), depending on the boundary localization given by $P_{\textrm{bound}}$. 
Experimental data points observed by DuChez et al. \cite{duchez2019durotaxis} are estimated to lie on the $Pe = 0.1$ line in the durotactic region (DT1), marked by the large stars. }
\label{fig:figure_FMI+DI}
\end{figure*}

So far, we have shown that elastic interactions  promote accumulation and trapping at the clamped boundary, thus facilitating durotaxis. On the other hand, cell motility enables escape from the boundary, thus counteracting durotaxis. We now quantify the extent of durotaxis in terms of some possible definitions of tactic index used in prior work. Based on our theory and simulations, we predict how the extent of durotaxis varies with the two main parameters in our model: the elastic cell-boundary interactions, $A=B$, and persistent cell motility, $Pe$. We focus on the case of a clamped boundary relevant for the cell located on the softer part of the substrate.

The elastic interaction parameter in our model, $A \sim P^{2}/E$, can be tuned by varying substrate stiffness, $E$. For a cell with fixed contractility $P$, the elastic interaction scales inversely with $E$, thus predicting a reduction in durotaxis with increasing substrate stiffness. We first compare our predictions with DuChez \emph{et al.} \cite{duchez2019durotaxis}, where the authors observed durotaxis of migrating U-87 glioblastoma cells up a stiffness gradient on polyacrylamide substrates. They quantified the extent of durotaxis as a forward migration index (FMI), defined as the ratio of the displacement of a cell up the stiffness gradient  to its total path length. In our simulation setup, this corresponds to  $-{\Delta}x/\ell_{p}$,  that is, the ratio of displacement of the cell towards the clamped boundary to the total path length traversed  along its trajectory. The substrate in the experiment comprised of  three, connected, $250$ \si{\mu m}-wide regions, labeled ``soft'', ``medium'', and ``stiff'', with average Young's moduli ($E$) of $5$ kPa, $10$ kPa and $15$ kPa, respectively. This allows us to map the dependence of a tactic index on  $A$ and $Pe$ and enables quantitative comparison of experimental observations with our model predictions. 

Using typical values for cell diameter, $\sigma \sim 20$ ${\mu}$m, and traction forces $\sim 2.5$ nN \cite{ketebo2021probing}, we estimate the elastic interaction parameter $A= B$ to be $5$, $2.5$, and $1.7$, corresponding to the three average substrate stiffness values in the experiment. We estimate $ Pe \sim 0.1$  for cells in all these regions, based on their measured migration speed, $ v_0\approx 0.4 \mu$ m/hr, and persistence time, $ D_R^{-1} \approx 0.1$ hr.  The results from the simulation are plotted along with experimental data in Fig. \ref{fig:figure_FMI+DI}(a). We find that  the three data points for FMI from the experiment agree closely with those obtained from simulations for corresponding estimated $A=B$ values. Overall, this demonstrates that durotaxis increases when the cell is initially on softer substrates.

To classify our simulated results into qualitatively different regimes, we define tactic indices that predict the dependence of durotaxis on two key model parameters. These are: $A$ (here we have chosen $B=A$), which represents the elastic cell-boundary interactions that drive durotaxis, and the persistent cell motility represented by $Pe$. Higher values of $Pe$ induce accumulation of cells at a confining boundary but also facilitate escape from ``durotactic trapping'' induced by the elastic potential.  Thus, in our model setup, accumulation does not imply durotaxis. To distinguish accumulation from durotaxis, we define and calculate a durotactic index (DI), that is distinct from the propensity to accumulate at a confining boundary given by $p_{\textrm{bound}}$. 
To define DI, we need to consider the accumulation driven by elastic interactions alone. We thus compare $N_{\textrm{bound}}$, the number of occurrences of a cell at the boundary at steady state, at some motility $Pe$, for $A\neq0$ and $A=0$:
\begin{equation}
\label{eq:DI}
    \mathrm{DI} = \frac{N_{\textrm{bound}}(A,Pe)-N_{\textrm{bound}}(A = 0,Pe)}{N_{\textrm{bound}}(A,Pe)+N_{\textrm{bound}}(A = 0,Pe)}.
\end{equation}
This definition allows us to subtract out the effect of motility-induced accumulation from the net accumulation. {This may be visualized in the simulation setup shown in Fig.~\ref{fig:figure_FMI+DI}(b)}. 
In one case, we consider a confining boundary with clamped elastic boundary condition corresponding to $A \neq 0$, while in the other, the confining boundary has no elastic interactions, $A=0$.  The difference in the number of accumulated cells between the two boundaries at steady state is then our chosen measure of durotaxis. This is analogous to the definition of DI used in previous works \cite{novikova2017persistence,yu2017phenomenological}: DI $ = (N_{f} - N_{r})/(N_{f}+N_{r})$, the normalized difference in the number of steps $N_{f}$ in a cell trajectory in the ``forward'' direction - that is, the direction up a stiffness gradient,  and the number of steps $N_r$ in the ``reverse'' (down the stiffness gradient) direction,. 

Next, we synthesize all simulation results for the clamped boundary case and organize them into a phase diagram in the space spanned by $A$ and $Pe$. In this simulated phase diagram shown in Fig.~\ref{fig:figure_FMI+DI}c, we classify the region corresponding to DI above a critical value (DI $\geq 0.27$)  to be ``durotactic''.  This choice corresponds to the calculated value of DI at $Pe = 0$, $A = 1$, since we expect elastic attraction to dominate over diffusive (random) cell motion for $A \geq 1$. The phase boundaries are constructed by interpolating  through 200 simulation data points ($A = 0$ to $10$ and $Pe = 0$ to $10$). The durotactic region can be further separated into two regimes by the line $Pe = A$. The $Pe < A$ region corresponds to a diffusion-dominated regime (DT1), where escape from the attractive boundary is facilitated by cell protrusion-facilitated random motion. The motility-dominated regime (DT2) occurs when $Pe > A$, and in this case escape from the attractive boundary is driven by persistent motility, without requiring any separate diffusive motion. Thus, in each case, it is the random or persistent motility, given by $D$ and $Pe$ respectively, which primarily competes with elastic interactions to reduce durotaxis.

For $A<1$ or at high motility relative to elastic interactions $Pe \geq 5 A$, the cells do not show sufficient durotaxis. These cells yield DI $< 0.27$, and are not considered to be in the DT regime. They can still accumulate at the boundary if the motility is high enough. We denote this latter regime ``motility induced accumulation'' (AD2-MIA), and distinguish it from the adurotactic (AD1) region without accumulation, using a threshold value of $P_{\textrm{bound}}$. At $A = 0$, we consider the value of $P_{\textrm{bound}}$ at $Pe = 5$ to be the cut-off value ($P_{\textrm{bound}}$ $= 0.18$) to separate regions AD1 and AD2(MIA). $P_{\textrm{bound}}$ $>0.18$ corresponds to MIA while $P_{\textrm{bound}}$ $<0.18$ corresponds to AD1.  All three datapoints from the DuChez et al. experiment \cite{duchez2019durotaxis} shown in  Fig.~\ref{fig:figure_FMI+DI}a lie in the DI region of the phase diagram and are indicated by large stars in the phase diagram in Fig.~\ref{fig:figure_FMI+DI}c. 

The main prediction of our simulated phase diagram is that durotaxis occurs when the strength of cell-boundary elastic interactions is large enough compared to random or persistent cell motility. This is realized when $A > A_{c}$, where the threshold value $A_{c}=1$ at $Pe=0$, and decreases with $Pe$. Higher values of $A$ can result from increased cell contractility, reduced substrate stiffness and/or less random cell movement. Higher persistent motility (larger $Pe$) helps the cell overcome the elastic boundary attraction and reduces durotaxis. While the predicted dependence on substrate stiffness is borne out  by the data from Ref.~\citenum{duchez2019durotaxis}, the dependence on migration speed ($Pe$) is yet to be systematically tested in experiments
because of the low value  $Pe < 1$ for cell migration in many cases.

\section{Discussion}
\begin{figure*}[ht]
\centering 
\includegraphics[width=0.91\linewidth] {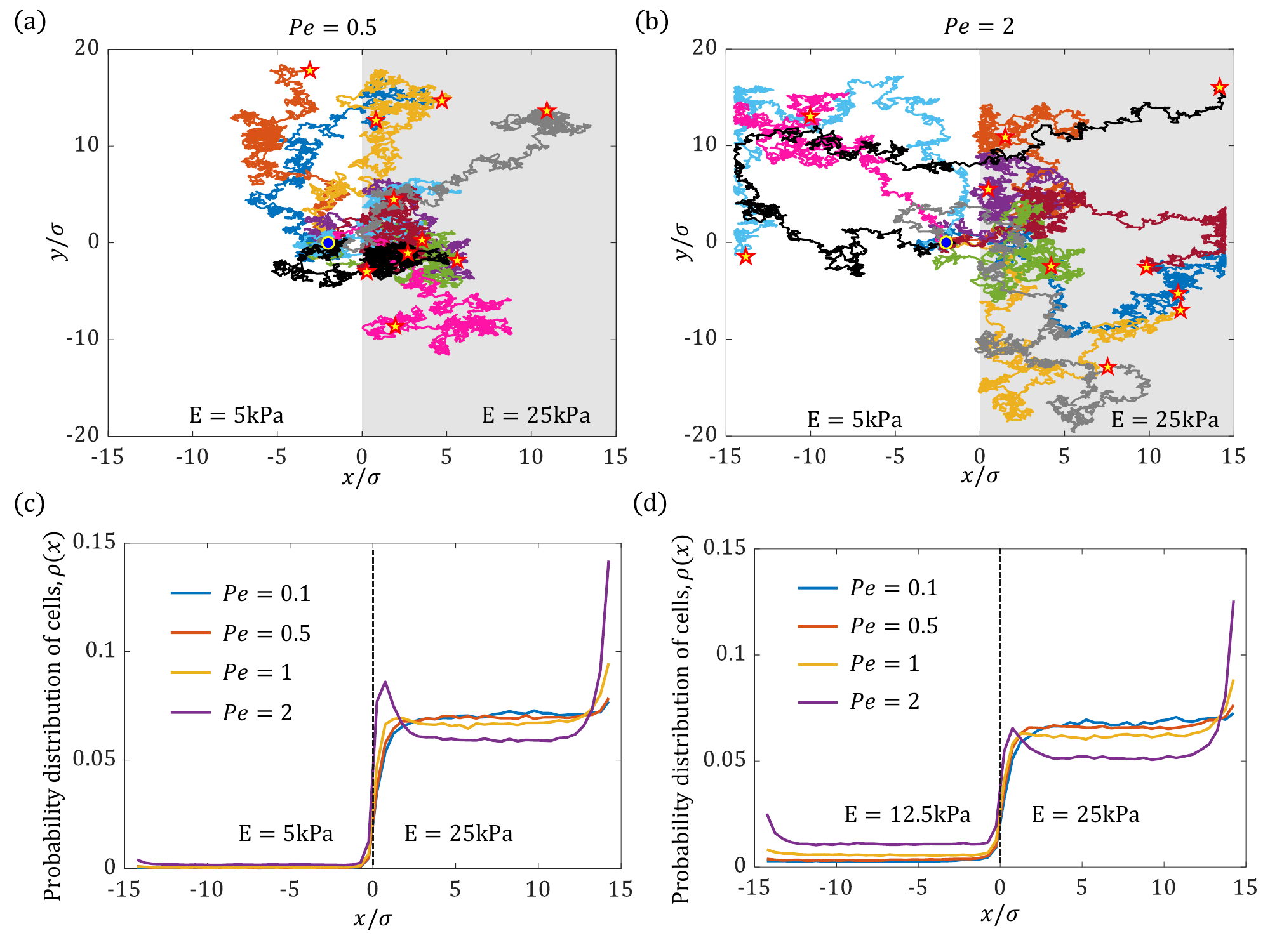}
\caption{\textbf{Durotaxis across sharp gradient of substrate stiffness} modeled by clamped and free boundary conditions - (a, b) Representative cell trajectories allowed to move across an interface between two regions (distinguished by white and gray) of contrasting substrate stiffness. In this example, they are chosen to have representative values of the Young's modulus of 5 kPa and 25 kPa, corresponding to $A = B = 5$ and $1$,  respectively. 
Each plot shows 10 single cell trajectories starting at $x/\sigma = -2$ with $D = 1$ (marked by a yellow disk) and terminating at different end points (marked by filled yellow pentagrams) after a total simulated time of $T = 20$. (a) All cells with lower $Pe = 0.5$ cross over to and spend more time on the stiffer side. (b) A few of the cell trajectories with $Pe = 2$ spend more time on the softer side as compared to the ones at lower values of $Pe$. (c, d) The steady state probability distribution demonstrates higher probability of finding cells on the stiffer side. The small probability of finding cells on the softer side is less for higher stiffness contrast in (c). It increases with decreased stiffness contrast in (d). The trend is more apparent at higher $Pe$, which allows cells to escape the attractive boundary force and spend more time on the softer side. Higher $Pe$ also lets the cells overcome the repulsion on the stiffer side, and form the small peak near the interface.}
\label{fig:figure_durotaxis_2sides}
\end{figure*}
  In this work, we combine a static elastic dipole model for cell-substrate mechanical interactions with a phenomenological model for persistent cell motility. We use this model to simulate cell dynamics and durotaxis at an elastic interface. The elastic dipole model for cell traction was invoked by Bischofs et al. \cite{Bischofs2003, Bischofs2004} to rationalize experimental observations of Lo et al. \cite{lo2000cell} that a fibroblast that is initially on the stiffer (softer) region, changes its orientation and aligns parallel (perpendicular) to the interface. The model as proposed was static without any cell dynamics, whereas we incorporate here both persistent and random contributions to cell motion. In this model setup, the accumulation of cells at the clamped (attractive) boundary facilitates durotaxis, since these cells can then cross over to the stiffer side.  On the other hand, the motility-assisted escape from this boundary reduces durotaxis, since the cell can reorient and make its way back to the softer side.  Our predictions for the reorientation (flipping) time given in Eq.~(\ref{eq:flip_time}) and cell migration index values (Fig.~\ref{fig:figure_FMI+DI}) may be used to infer how durotaxis depends on cell traction force (via $A$, and $B$), substrate stiffness values (also via $A$ and $B$), and motility (via $Pe$).  

Based on our simulations, we predict a phase diagram of cell durotactic behavior. We show that durotaxis is enhanced when the cell-substrate elastic interactions are large enough (high $A=B$), and the cell is not very persistently motile (low $Pe$). Our results quantitatively explain the finding by DuChez et al. \cite{duchez2019durotaxis} that the tactic index decreases with increasing local substrate stiffness.  Our results are also qualitatively supported by the recent observation of Yeoman \emph{et al.} that weakly adherent breast cancer cells show comparatively less durotaxis than their strongly adherent counterparts ~\cite{yeoman2021adhesion}. Weakly adherent cells are expected to undergo rapid assembly/disassembly of focal adhesions leading to faster motility as was indeed observed in the study. Faster cells are expected to have higher $Pe$ value according to an  established universal exponential correlation between cell migration speed and persistence \cite{Maiuri2015} based on experimental data. The observation that breast cancer cells are less durotactic is thus consistent with our predicted inverse relationship of durotaxis and persistent motility, seen in the phase diagram in Fig.~\ref{fig:figure_FMI+DI}c.

Yeoman \emph{et al.} performed traction force measurements and drug-treatment assays that inhibit the actomyosin cytoskeletal activity, but did not separately measure the effects of drug treatment on cell motility and contractility.  
Further experimental exploration using substrates of varying stiffness and adhesivity (e.g. by micropatterning) is needed for quantitative and conclusive comparisons with our theoretical predictions for the dependence of durotactic index on cell traction and migration velocity. We also predict a motility-induced accumulation regime where cells are expected to be preferentially located near a confining boundary. While this has been demonstrated for active synthetic particles and swimming bacteria, elucidating this hitherto unexplored effect for crawling cells requires experiments on micropatterned substrates. Future experiments can also test our model prediction that a cell can detect and respond to a sharp interface in  substrate stiffness from a long range (a distance of a few cell lengths away), without needing to be in direct contact with both softer and stiffer regions of the substrate.

To directly demonstrate durotaxis in our model, we consider the movement of cells across a sharp interface between two regions with contrasting substrate stiffness. In this simulation setup shown in Fig.~\ref{fig:figure_durotaxis_2sides}, the left side has a lower stiffness than the right side of the interface. The left and right boundaries at $x/\sigma = \pm 15$ provide only confinement and not elastic interaction.  We use the simplifying assumption of large stiffness contrast, such that a cell in the $x<0$ ($x>0$) region is considered to be interacting with a clamped (free) boundary, respectively. In Figs.~\ref{fig:figure_durotaxis_2sides} a and b, we show representative trajectories of single cells initialized on the softer side and close to the interface.  Most cells are seen to cross over to the stiffer side, but for higher $Pe$ values, a few are able to make their way back to the stiffer side. This illustrates our central point: that persistent motility can compete with elastic interactions. The steady state probability distributions in Figs.~\ref{fig:figure_durotaxis_2sides}c and d  further illustrate that a lower stiffness contrast leads to lower durotactic index. This is especially apparent at higher $Pe$, when the difference in number of cells between the two regions is reduced for lower stiffness contrast. Further, the higher $Pe=2$ cells show some motility-induced accumulation on the repulsive side of the interface, whereas at lower $Pe < 2$, a depleted layer is seen as the self-propulsion is unable to overcome the repulsion.

We use the approximately clamped or free boundary condition limits because the general elastic interaction potential between two substrate regions with arbitrary stiffness values lacks a simple analytically tractable form  \cite{walpole1996elastic}.  Further, when a cell moves across the stiffness interface, other shorter-range effects beyond the scope of this study are expected to dominate its dynamics. For example, a cell that can extend across the interface will deform the soft side more than the stiff side, leading to an effective translation towards the latter, which may drive durotaxis across gradual gradients in stiffness \cite{sunyer2020durotaxis}. Analogously, short-range effects are thought to drive ``viscotaxis'' of microswimmers \cite{datt2019, stehnach2021viscophobic}, in addition to longer range hydrodynamic interactions with an interface \cite{shaik2021hydrodynamics}. In this latter context, scattering or change in direction of microswimmers, analogous to refraction of light, has been seen to occur across a viscosity interface \cite{coppola2021green}. In the SI Fig. S4, we consider such effects in the zero noise ($D=0$), limit of our model, and show that  a scattering close to the interface also results from the elastic potential.

Recent observations of ``negative durotaxis'' or ``anti-durotaxis", \emph{i.e.,} directed migration from softer to stiffer substrates suggest that cells do not always move \emph{up} stiffness gradients, but rather move towards an optimal substrate stiffness where their contractility is maximal \cite{isomursu2022directed}. We note that the elastic dipole model can give rise to such an optimal stiffness when the mechanosensitivity of the cell to substrate properties is incorporated by including explicit feedback between cell traction force (the contractile dipole strength) and substrate deformation \cite{zemel_10}. This is motivated by experiments that suggest that cells sense and adapt their traction and effective force dipole moment to substrate strain \cite{Ghibaudo2008}. The inclusion of cell polarizability in the elastic dipole model creates additional interaction terms of the cell dipole with its image dipoles induced by the confining boundary. These additional pairwise interaction terms can be stronger and have the opposite sign from the direct interactions \cite{assi}. This may result in the clamped (free) boundary switching roles and being repulsive (attractive), which would drive negative durotaxis in our model. 
Alternatively, some adherent cells are known to be capable of regulating their traction forces to maintain different types of mechanical homeostasis depending on substrate stiffness~\cite{Wolfenson2020,Sam2015}. In the derivation of the (attractive) cell-boundary interaction energy used in this work in Eq.~(\ref{eq:Potential_full}), cellular forces (or dipole moment) have been assumed to be constant, indicating stress or force homeostasis. If, instead, cells maintain constant displacement (known as displacement homeostasis)~\cite{Wolfenson2020}, then the attraction to the rigid boundary could turn repulsive~\cite{Sam2015}, resulting in negative durotaxis.
These effects will be explored in future work. In general, our work paves the way for exploring active cell migration under confinement and various tactic stimuli \cite{yu2022biphasic} that may be expressed as effective potentials. 


%



 
\section*{Acknowledgements}
SB and KD acknowledge support from the National Science Foundation (NSF-CMMI-2138672) and also support from NSF-CREST: Center for Cellular and Biomolecular Machines (CCBM) at the University of California, Merced via NSF-HRD-1547848. HW and XX are supported by the National Natural Science Foundation of China (NSFC, No. 12004082, No. 12374209). We acknowledge useful discussions with Assaf Zemel, Yariv Kafri, Samuel Safran, Alison Patteson, and Ajay Gopinathan.

\section*{Appendix}
\appendix
\section{Model for substrate mediated cell-interface interactions}
Adherent cells exert dipolar contractile stresses on the underlying elastic substrate\cite{oakes2014geometry}; these are generated by actomyosin fibers (actin and myosin II complexes), usually referred to as stress fibers\cite{Dasbiswas2018}, that generally connect the opposite sides of the cell and terminate at focal adhesions (FAs)~\cite{Mohammadi2014,Sam2013a,Murrell2015}. On a larger scale, the entire contractile cell can be represented as a force dipole that deforms its extracellular environment typically modeled as a linear elastic continuum \cite{Sam2013a, Schwarz2001}. The concept of force dipoles has found wide-ranging applications in various biological phenomena.~\cite{Bischofs2003,Bischofs2004, Dasanayake2011,Soares2011,Sam2013a,De2007dynamics,zemel_10,Ranft2010fluidization}.

Here, we use the force dipole model and extend current theory to the interactions of active, motile cells with an underlying elastic substrate and constrained to remain within a domain (with boundaries) using a combination of simulations and analytical theory.  In this minimal model, the entire, polarized cell, is coarse-grained and approximated as a single, evolving force dipole that moves on an elastic substrate, and is further subject to forces generated due to its interaction with the substrate and its boundaries. For the purposes of the analysis however, we use the word active to specifically mean self-propelling cells. 
Given the assumption of isotropic linear elasticity of the extracellular material, and the strength and orientation of the cell generated dipole, we can calculate stress and strain fields by solving the elastic equations with appropriate boundary conditions. These stress/strain fields then affect the motion of the cell by allowing cells to re-orient towards preferred alignments in order to optimize the deformation energy generated by the dipole in the substrate. Two canonical reference cases, namely 1) free boundaries, where the normal traction vanishes at the stiff-soft boundary (useful to analyze cells located on stiffer side), and 2) clamped
boundaries, where the displacements vanish at the stiff-soft boundary (relevant to cells initially located on softer side) are analyzed. Such reduced descriptions are particularly appropriate when the stiffness contrast is high.  The corresponding elastic boundary value problems with these limiting boundary conditions can be solved using the method of images~\cite{Bischofs2004}.  

In general, the interaction energy of the adherent cell (force dipole) with the surface \cite{Bischofs2004} scales as  $U\sim P^2f_{\nu}(\theta)/(E x^3)$, where $f_{\nu}$ is a function of substrate Poisson's ratio $\nu$, and the orientation of the cell relative to boundaries. 
Here, the spatial and angular coordinates $x$ and $\theta$ are as defined in Fig. 1 -(c). 
The substrate mediated elastic cell-boundary interaction can be modeled as an effective potential $U(x,\theta)$ acting on the adherent cells (generating a force dipole) thus, 
\begin{equation}
\begin{aligned}
U(x, \theta) &= -\left(\frac{P^2 }{256\pi E}\right)\frac{f_{\nu}(\theta)}{\left(x + \sigma \right)^3},\nonumber \\
f_{\nu}(\theta) &= a_{\nu}+b_{\nu}\cos ^2 \theta+c_{\nu} \cos ^4 \theta,
\end{aligned}  
\end{equation}
with $P$ being the force dipole, $E$ and $\nu$ being the Young's modulus and Poisson's ratio of the substrate, respectively. The parameters $a_{\nu}, b_{\nu}, c_{\nu}$ are different for free and for clamped boundary conditions. These are, respectively (with superscript $f$ denoting free, and superscript $c$ denoting clamped)
\begin{equation}
\begin{aligned}
a_{\nu}^f & =-\frac{(1+v)[5+2 v(6 v-1)]}{(1-v)},\\ 
a_{\nu}^c & =\frac{(1+v)[15+32 v(v-1)]}{(1-v)(3-4 v)} \\
b_{\nu}^f & =-\frac{(1+v)[22+4 v(2 v-9)]}{(1-v)},\\ 
b_{\nu}^c & =\frac{\left.(1+v)\left(34+32 v^2-72 v\right)\right]}{(1-v)(3-4 v)} \\
c_{\nu}^f & =-\frac{(1+v)\left[13(1-2 v)+12 v^2\right]}{(1-v)}, \\ 
c_{\nu}^c & =\frac{(1+v)(7-8 v)}{(1-v)(3-4 v)}
\end{aligned}
\end{equation}
Preferred cell orientations, as predicted by calculating configurations that minimize deformation energy, are parallel/perpendicular to the boundary line for free/clamped boundaries. Hypothesizing that this holds even for motile cells, and accounting for the effects of self-propulsion, we deduce that motile cells preferentially move toward a clamped boundary, but tend to migrate away from a free boundary.

In addition to elastic effects, boundaries may physically constrain cells from crossing. This constraint is implemented by explicit displacements of the cells, as explained in the next section.

\begin{table} [ht]
\caption{List of biophysical parameters.\label{tab1}}
\begin{center}
\begin{tabular}{lll}
\hline
\textbf{Parameter}	& \textbf{Meaning}	& \textbf{Value(s)}\\
\hline
$\sigma$		& Cell diameter	& $10 - 100$ \:\:$\mu$m       \\
$v_{0}$		& Cell velocity	& $0 - 80$ \:\:$\mu$m hr$^{-1}$       \\
$\mu_{T}$	& Translational Mobility & 0.1\:\: m$^2$ min$^{-1}$ pN$^{-1}$        \\
$\mu_{R}$		& Rotational Mobility			& $25$\:\:$\mu$m$^2$ min$^{-1}$        \\
$D_{\mathrm{eff}}$		& Rotational Diffusivity			& $0.01 - 0.1$\:\:min$^{-1}$        \\
$E$		& Young's modulus			& $0.5 - 100$ kPa     \\
$\nu$		& Poisson's ratio			& $0.3$    \\
$P$ &  Contractility & $10^{-12} - 10^{-11}$ \si{N \cdot m}  \\
\hline
\end{tabular}
\end{center}
\end{table}

\begin{table*} [ht]
\caption{List of simulation parameters.\label{tab2}}
\begin{center}
\begin{tabular}{llll}
\hline
\textbf{Parameter}	& \textbf{Meaning}	& \textbf{Definition} & \textbf{Value(s)}\\
\hline
$A$ & Cell-boundary force parameter & $\frac{\mu_{T}P^{2}} {(E D_{T} \sigma^{3})}$ & $0.1 - 100$    \\
$B$ & Cell-boundary torque parameter & $\frac{\mu_{R}P^{2}} {(E D_{R} \sigma^{3})}$ & $0.1 - 100$    \\
$Pe$ & P{\'e}clet Number  & $\frac{v_{\mathrm{0}}} {\left(\sigma D_{\mathrm{R}}\right)}$ & $0 - 10$    \\
\hline
\end{tabular}
\end{center}
\end{table*}

\section{Simulation model details}

The position and orientation of the cells is governed by over-damped Langevin equations. The simulation box has a square geometry with lateral dimension $L$ with $x$ representing the scaled distance measured normal to the boundary (see Fig. 1).  We perform the simulations in dimensionless units. To do this, we choose $1/D_{R}$ as the characteristic time scale, and introduce dimensionless time $t^{*}$ related to dimensional time $t'$ by $t^{*} \equiv t' D_{R}$. The diameter of the cell $\sigma$ is used to scale lengths, so that the dimensionless positions $(x^{*},y^{*})$ are related to the dimensional ones $(x', y')$ via $x^{*} \equiv x'/\sigma$, $y^{*} \equiv y'/\sigma$ and ${\bf {r}}^{*} \equiv {\bf r}'/\sigma$.  The equations when scaled assume the form
\begin{eqnarray}
    \frac{d\mathbf{r}^*}{dt^*} &=& {Pe}\:\mathbf{p}\: - \frac{3A}{(x^{*}+1)^4} \Tilde{f}_{\nu}(\theta)\hat{\mathbf{x}} + \sqrt{2D}{\bm{\eta}^*_{T}}
    \label{eq:Model_ND_A} \\
    \frac{d\theta}{dt^*} &=& -\frac{B}{(x^{*} + 1)^3} \frac{\partial \Tilde{f}_{\nu}(\theta)}{\partial \theta} + \sqrt{2} \eta^*_{R}
    \label{eq:Model_ND_B}
\end{eqnarray}
where $A$ and $B$ are the dimensionless interaction parameters for force and torque respectively and $Pe$ is the P{\'e}clet number which determines the persistent motion of the cells(Eq.~\ref{eq:A_B_Pe}). $D$ is the scaled coefficient of diffusion (Eq.~\ref{eq:A_B_Pe}) while $\bm{\eta}_{T}^*$ and $\eta_{R}^*$ are the scaled Gaussian white noise for translation and rotation respectively. In our simulations $\nu$ is fixed at $0.3$ \cite{lo2000cell} and $f_{\nu}(\theta)$ is scaled such that $\Tilde{f}_{\nu}(\theta) = (50/256\pi) f_{\nu}(\theta)$. 
Superscripts $^{*}$ in Eqs.~(\ref{eq:Model_ND_A}) and (\ref{eq:Model_ND_B}) denote non-dimensional quantities. Henceforth, we will drop this superscript for ease of use and thus in the final equations simulated $(x,y,t)$ are all dimensionless.



%


Simulations are conducted, unless mentioned otherwise,  with $N=200$ active Brownian ps (cells) of diameter $\sigma$. In scaled units, the cells have diameter of 1, and move within a square box of size $L = 40$. Cells do not interact with each other. We choose the origin and coordinate axes $x$ and $y$ so that the domain is $-L/2 \leq x \leq L/2$ and $-L/2 \leq y \leq L/2$.  Periodic boundary conditions are imposed at the lower and upper boundaries.  

Lateral boundaries correspond to flat interfaces that interact elastically with cells and also impose confinement. We ignore deformations of the boundary so that these interfaces are always parallel to the y-axis at $x = -L/2$ and $x = L/2$. Confinement is directly imposed by maintaining an exclusion region of $\sigma/2$ exists  around each interface; cells are thus prevented from partially or fully penetrating the wall. We implement this condition as follows. We make sure that if a particle makes a virtual displacement where the center of the particle is $x + \Delta x > L/2 - \sigma/2$, it is brought back to a distance $L/2 - \sigma/2$ and similarly to $-L/2 + \sigma/2$ on the other confining boundary. The free and clamped boundary conditions are associated with the confining boundaries to ensure that the particles cannot cross the threshold potential. The coordinate system shown in Fig. 1-(c), demonstrates symmetry (in both the type of boundary conditions, and potential field from the boundary) about the origin $x = 0$, and reflection symmetry about the $y$ axis. Since $x$ denotes the  variable quantifying the normal distance measured  from the edge of the boundary, our simulation methodology implies that particles are excluded from occupying a region of width $1/2$ (corresponding to the radius of the cell $\sigma/2$ in dimensional units) at the boundary (see Fig. 1-(c, f)).

Dimensionless forms of the dynamical equations Eq.~\ref{eq:Model_ND_A}-\ref{eq:Model_ND_B} are discretized and numerically solved using the explicit half-order Euler-Maruyama method \cite{Allen2017}. We initialize 200 non-interacting particles uniformly distributed inside the simulation box and study its probability distribution as function of distance from the boundary. These particles interact with the elastic boundaries depending on the proximity and orientation with respect to the boundary. Simulating a large number of non-interacting cells at the same time allows us to obtain detailed statistics for single particle interaction with the elastic boundary in a speedy and efficient manner. The dimensionless time step is $dt = 10^{-3}$ such that the displacement in each time step is small ($\sim 10^{-2}\sigma$ or smaller). We sample the data every $10^3$ steps. When the probability distribution does not change with time (subject to a pre-specified precision), we consider that statistical steady state has been reached. Steady state is achieved at different times which depend on the parameters $A$, $B$ and $Pe$. Steady state time under no force or torque from the boundary can be estimated to be $\sim L^2/D_{\mathrm{eff}}$ where $D_{\mathrm{eff}} = v_{0}^2/D_{R}$. In our initial simulations, we set the scaled translational diffusion $D = 0$. Thereafter, we study the distribution of particles as a function of distance from the boundary by averaging over all particles and time after steady state is achieved. We count the number of particles at $x = \sigma/2$ to determine the localization of particles at the boundary.

At steady state we look at the distribution of particles throughout the domain from the left wall to the midpoint of the domain, and also analyze the localization of particles near the boundary (over a region ranging from  a cell diameter to a few cell diameters). This is done by studying the time evolution of the effective number of particles/cells  a certain distance from the wall. If the interface was a penetrable surface, higher localization at the boundary would imply  a higher probability of cells and a larger current/flux crossing the interface. For a free boundary, we study the effect of simulation parameters  on the void length and orientation dynamics of particles at the clamped boundary. Our simulations complemented by a simple model for barrier crossing based on Kramer's theories allow us to identify conditions particles can escape the influence of the boundary interactions.

\begin{figure*}[hbt!]
\centering
 \includegraphics[width=0.8\linewidth]{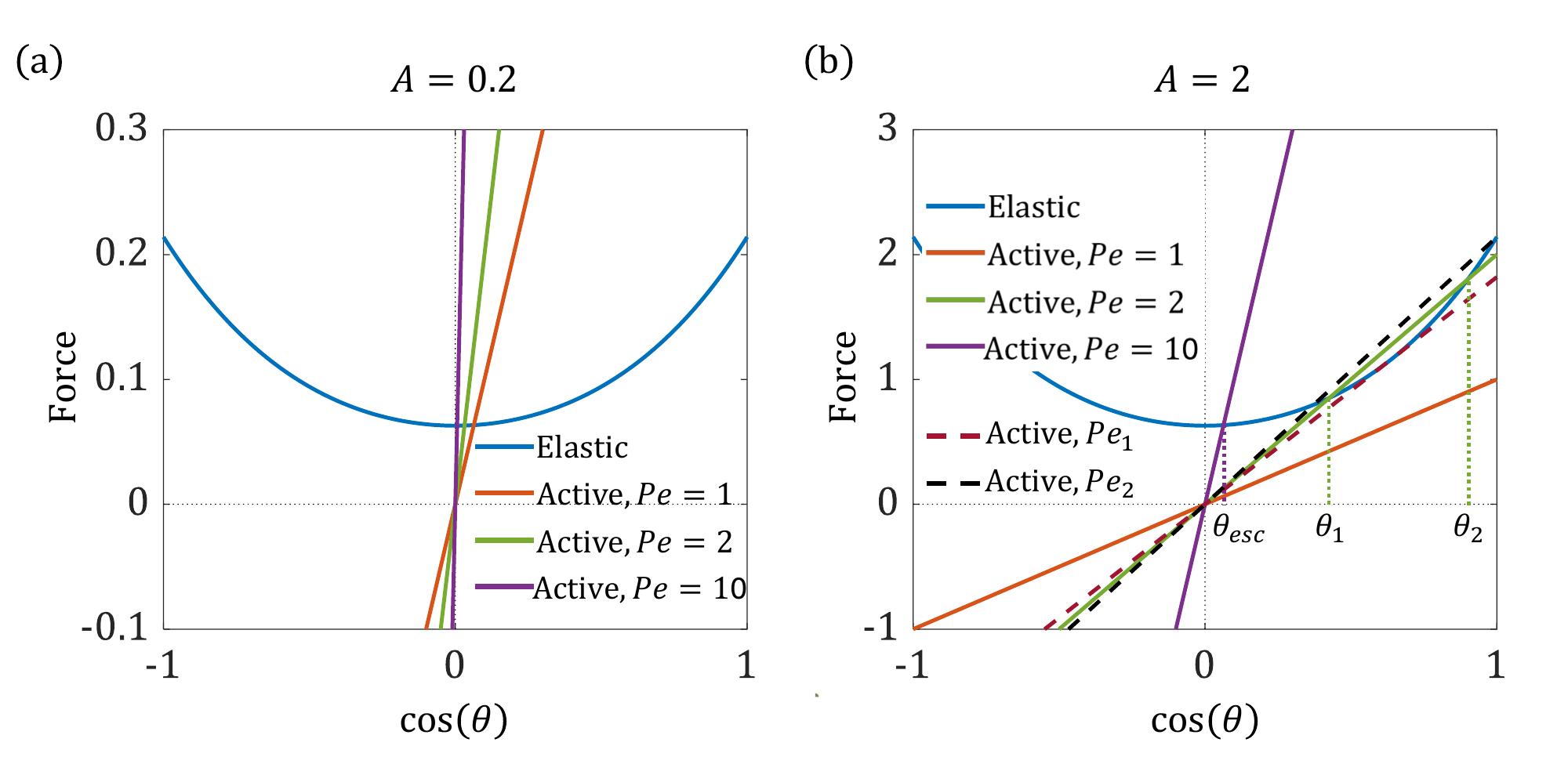}
\caption{The elastic force at the boundary and active force perpendicular to the boundary both depend on the angle of orientation with the boundary.
We compare the forces at the boundary to graphically estimate the angle of escape of the particles. We compare the force from the boundary (solid blue) at (a) $A = 0.2$ and (b) $A = 2$ with active forces  perpendicular to the boundary (dashed) at $Pe = 1, 2, 10$. The particle can escape at angles where the perpendicular component of the active force is greater than the boundary interaction. (a) At $A = 0.2$, for all values of $Pe$, the particles can escape the boundary through any angle $\theta$ such that $cos{\theta} > 0$. Increasing $Pe$ increases the angle of escape. (b) At $A = 2$ we observe 3 different behaviors. For $Pe = 1$ perpendicular component of active force is always less than the boundary force. At 
$Pe = 10$, the active force is higher than the boundary force and intersect each other at 1 point. The active force is higher than the boundary force only inside angular pockets for $Pe = 2$.}
\label{fig:Escape-angle}
\end{figure*}

\section{Determining escape condition for cells trapped at clamped boundary}

Here we graphically determine the criterion for escape of particles from the boundary at different values of the interaction parameter, $A$, and P{\'e}clet number, $Pe$. We further determine the critical values $Pe_{1}(A)$ and $Pe_{2}(A)$ which dictate the different regimes of particle localization at the boundary. Particles remain trapped at the boundary when $Pe < Pe_{1}$. For $Pe > Pe_{2}$, there exists a characteristic angle $\theta$, above which trapped particles can attain a configuration favorable for escape from the boundary. This critical angle, $\theta_{\textrm{esc}}$ depends on $Pe$ (Fig.5(a), Fig.~\ref{fig:Escape-angle}(a)). For $Pe_{1} < Pe < Pe_{2}$ particles can only escape the boundary when their orientation $\theta$ lie in the angular region between $\theta_{1}$ and $\theta_{2}$ (Fig. 5(b)).
%

The particles can escape when the self-propulsive active force of the particle has an perpendicular component sufficiently large to overcome the elastic attraction from the boundary. For a particle/cell trapped at the boundary $x=x_{b} =\sigma/2$, force balance yields 
\begin{equation}
\begin{aligned}
Pe  = { 1 \over \cos{\theta}} \: \frac{3A}{\left(x + 1\right)^4}\Tilde{f}_{\nu}(\theta), \quad \Tilde{f}_{\nu}(\theta) = \frac{50}{256\pi}f_{\nu}(\theta).
\label{eq:force_bal_boundary_app}
\end{aligned}
\end{equation}
At $Pe = Pe_{1}$, the tangent construction evaluating the elastic force due to cell-boundary interactions (see Fig.~\ref{fig:Escape-angle}) gives $Pe_{1}$,
\begin{equation}
    Pe_{1} = \frac{3A}{(x+1)^4}\frac{50}{256\pi}\left(2|b_{\nu}^{c}|\cos{\theta} + 4|c_{\nu}^{c}|\cos^3{\theta}\right)
\label{eq:Pe1_app}
\end{equation}
Eqs.~(\ref{eq:force_bal_boundary_app}) and (\ref{eq:Pe1_app}) provide the ratio $Pe_{1}/A = 0.91$ at $x = x_{b} = \frac{\sigma}{2}$. At $A = 2$, $Pe_{1} = 1.819$ and at $A = 0.2$, $Pe_{1}$ is expected to be $0.182$.\\
To determine $Pe_{2}$, we consider $\theta_{1} = 0$, since beyond $Pe_{2}$, $\theta_{1}$ would cease to exist as particles can escape at angles less than $\theta_{2}$. Balancing forces at $\theta = 0$, we get
\begin{equation}
    Pe_{2} = \frac{3A}{(x+1)^4}\frac{50}{256\pi}(|a_{\nu}^{c}| + |b_{\nu}^{c}| + |c_{\nu}^{c}|)
\label{eq:Pe2_app}
\end{equation}
This gives the ratio $Pe_{2}/A = 1.07$. For $A = 2$, $Pe_{2}$ is determined to be $2.14$ and for $A = 0.2$, it is $0.214$.


\section{Kramer's theory for the frequency of orientation flips for spatially localized cells}


We analyze the flips in cell orientation, that is in the angle $\theta$,  when the cell is at a fixed location near the boundary. This is done via an adaptation of the classical theory due to Kramer~\cite{Kramers1940}. Consider a collection of independent Brownian cells/particles in an external 1D potential $U(z)$ that depends on a generalized coordinate $z$. 
The well is sufficiently deep so particles inside it cannot escape at short time intervals.  Assuming that particles in the well minima are close to equilibrium and cross the barrier diffusively, we aim to obtain the rate at which this escape takes place. 
The dynamics of a test particle can be described by the over-damped Langevin equation in 1D,
\begin{equation}
\frac{d z_{p}}{d t}=-\mu U^{\prime}(z_{p})+\eta(t)
\end{equation}
with $\mu$ being the mobility and $-U^{\prime}(z_{p})$ the linear drag force acting on the particle located at $z_{p}$. The particle is also subject to a white noise $\eta(t)$, with zero mean $\langle\eta(t)\rangle=0$ and variance $\left\langle\eta(t) \eta\left(t^{\prime}\right)\right\rangle=2 D \delta\left(t-t^{\prime}\right)$. Here $D$ and $\mu$ are generalized diffusivity and mobility coefficients that characterize the random diffusion and frictional effects as the particle moves along $z$.  Barrier crossing is achieved after many attempts, that is, the crossing is driven by diffusive processes. 

These approximations allow us to move from the Langevin equation to the Fokker-Planck equivalent. We recast the problem in terms of a probability distribution function $P(z,t)$ that may be mapped to either the probability of a single particle or the density of a collection of non-interacting particles. We assume that the system is close to equilibrium so that crossing flux $J(z,t)$ may be related to gradients in  $P(z,t)$,
\begin{eqnarray}
\label{eq:Fokker-Planck}
\frac{\partial P(z, t)}{\partial t} 
&=&\frac{\partial}{\partial z}\left[\mu \frac{\partial U}{\partial z} P +D \frac{\partial P }{\partial z}\right]
=-\frac{\partial J}{\partial z} \\
J &=&-\mu P(z,t)\partial_{z} U -D \partial_{z}P(z,t) \\
J &=&-D e^{-\frac{U(z)}{k_B T}} \frac{\partial}{\partial z}\left(e^{\frac{U(z)}{k_B T}} P(z)\right).
\end{eqnarray}
We have invoked the Stokes-Einstein relationship so that $D=\mu k_B T$. 
For a system that is approximately in equilibrium and in quasi-steady conditions with a large barrier height satisfying $[U(B)-U(A)]/(k_B T) \ll 1$, the current $J$ across the barrier is small
and the rate of depletion in the well is small.  Since the system is close to quasi-steady state, the probability distribution $P(z,t)$ does not change quickly with time, and so $\partial_t P(z,t)\approx 0$. Moreover, based on Eq.~\ref{eq:Fokker-Planck}, the current is then to leading order constant and  independent of $z$ and $t$. 
 \begin{equation}
 \label{eq:Current-Integrate}
    \frac{J}{D} e^{\frac{U(z)}{k_B T}} =-\frac{\partial}{\partial z}\left[e^{\frac{U(z)}{k_B T}} P(z, t)\right].
 \end{equation}

Due to the barrier crossing event being a rare event, we next invoke the approximation $P(A) \gg P(C)\sim 0 $. 
To calculate the escape flux, we assume that re-crossings into the well are not permitted once the particle reaches location $C$. That is, we let $C$ correspond to an absorbing boundary so that the probability density there is zero. Integrating Eq.~\ref{eq:Current-Integrate} between locations $A$ and $C$, and using $P(C) = 0$, we obtain 
\begin{equation}\label{eq:current-integral2}
 \frac{J}{D} {\int_A^C e^{\frac{U\left(z\right)}{k_B T}} dz}
 = e^{\frac{U(A)}{k_B T}} P(A).
\end{equation}
\begin{figure*}[ht]
\centering
\includegraphics[width=0.81\linewidth]{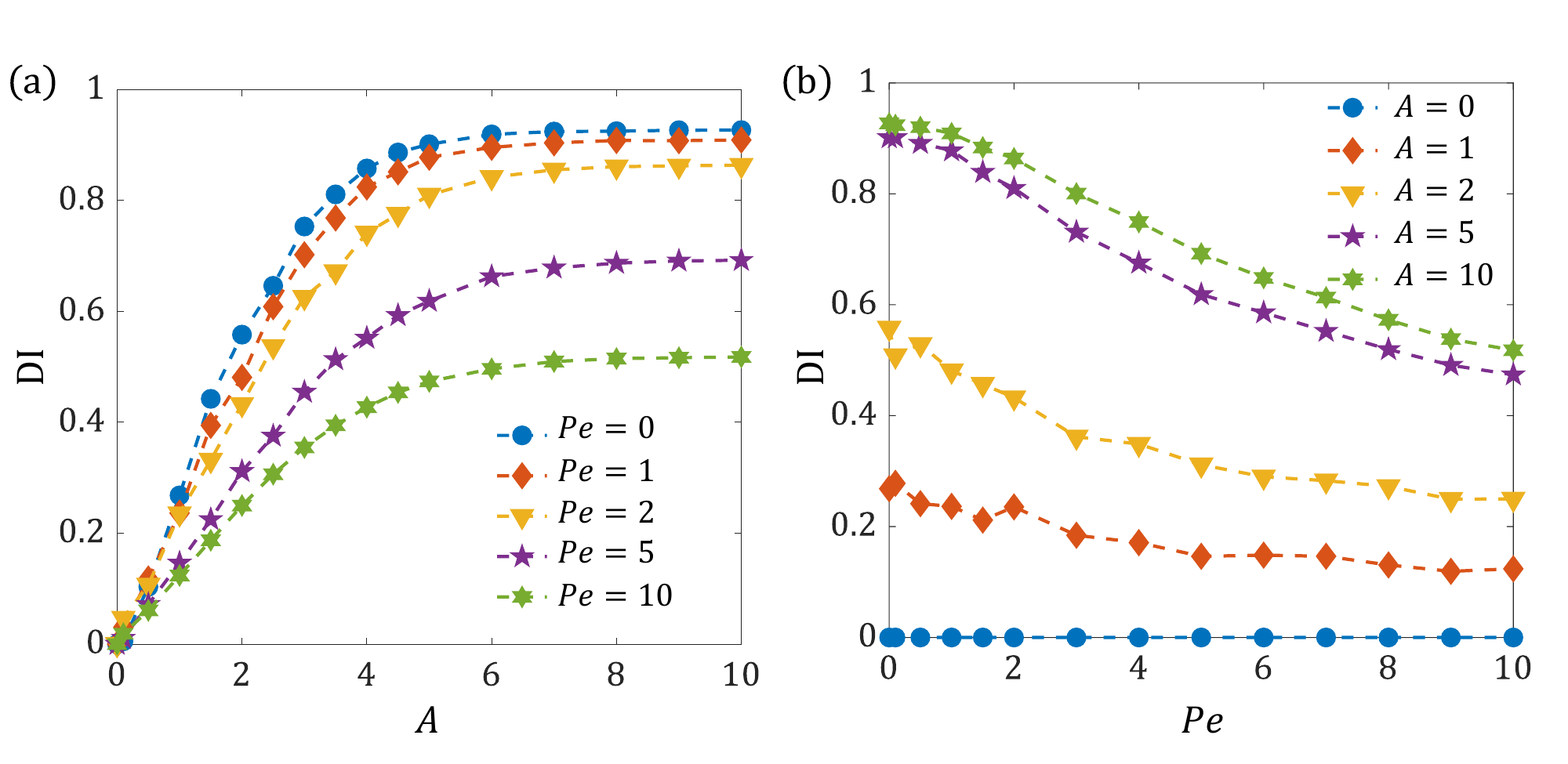}
\caption{The durotactic index DI as defined in Eq.~\ref{eq:DI} is plotted here in two ways. In (a), we observe that DI increases with elastic force parameter, $A$ when $Pe$ is held constant and reaches limiting values. (b) For $A$ held fixed, we find that DI decreases with cell motility, $Pe$. The index is $0$ by definition for $A=0$.}
\label{fig:DI_vs_A_and_Pe}
\end{figure*}
The left side integral can be asymptotically estimated to leading order by using the saddle point method by expanding $U(z)$ in a Taylor series approximation and noting that the first derivative at $B$ is zero,
\begin{equation}
\begin{aligned}
&\int_A^C e^{\frac{U\left(z\right)}{k_B T}} dz
\approx
\int_{A}^{C} e^{\frac{U(B)+\frac{1}{2} U^{\prime \prime}(B)\left(z-B\right)^2}{k_B T}} dz \\
&\approx
e^{\frac{U(B)}{k_B T}} \int_{-\infty}^{\infty} e^{\frac{-\left|U^{\prime \prime}(B)\right|\left(z-B\right)^2}{2 k_B T}} dz
=e^{\frac{U(B)}{k_B T}} \sqrt{\frac{2 \pi k_B T}{\left|U^{\prime \prime}(B)\right|}}.
\end{aligned}
\end{equation}
To evaluate the escape rate $r_{\mathrm{esc}}$, we recognize that this rate is the same as the current going out of the metastable well at $A$, given that the particles are initially situated inside it, $J=p_A r_{\mathrm{esc}}$. Assuming an initial close equilibrium state with 
\[
P(z)=P(A) \exp\left[-[U(z)-U(A)]/k_B T\right]
\] 
and using the expansion 
\[
U(z) \approx U(A)+\frac{1}{2} U^{\prime \prime}(A)(z-A)^2,
\]
the probability to be inside the well is approximately
\begin{equation}
\begin{aligned}
p_A =\int_{A-\Delta}^{A+\Delta} P(z)\: dz
& \approx P(A)
\int_{-\infty}^{\infty} e^{-\frac{U^{\prime \prime}(A)\left(z-A\right)^2}{2 k_B T}} dz \\
&=P(A) \sqrt{\frac{2 \pi k_B T}{U^{\prime \prime}(A)}}.     
\end{aligned}
\end{equation}
 Here $\Delta$ denotes a suitably small range in the neighborhood of point A. The saddle point approximation allows us to eventually extend  the domain of integration from $-\infty$ to $\infty$.
Thus, the escape time satisfies,
\begin{equation}
\tau_{\mathrm{esc}}
={1 \over r_{\mathrm{esc}}}
=\frac{p_A}{J}
={{2 \pi k_B T \exp{\left(\frac{U(B)-U(A)}{k_B T}\right)}} \over {D 
{\sqrt{U^{\prime \prime}(A)\left|U^{\prime \prime}(B)\right|}}}}
\label{eq:flip_time_app}
\end{equation}
To use Eq.~\ref{eq:flip_time_app} to study flipping dynamics, we consider a particle located at a {\em fixed position} $x$ and study the time it takes to reorient from $\theta=0$ (bottom of the potential well),  to $\theta = \pi/2$ (top of the barrier). 
The escape time can be mapped into a $1D$ circular motion with periodic boundary condition $P(\theta=0)=P(\theta=2\pi)$ \cite{sommerfeld1949partial}. Identifying the coordinate $z$ as $\theta$ and reintroducing the location dependence $x$ (here considered constant), we obtain the escape time at fixed $x$
\begin{equation}
\begin{aligned}
\tau_{\mathrm{esc}} D_{R}
&=\frac{2\pi k_B T \exp \left(\frac{U(x,\pi/2)-U(x,0)}{k_B T}\right) }{\sqrt{\left[ U^{\prime \prime}(x, 0) \left|U^{\prime \prime}(x, \pi/2)\right|\right]}}\\
&\sim \frac{1}{B} \exp \left(\frac{B}{B^*}\right).
\end{aligned}
\end{equation}

 \section{Measurement of tactic indices from simulation}
 
 \begin{figure*}[htbp!]
\centering
\includegraphics[width=0.61\linewidth]{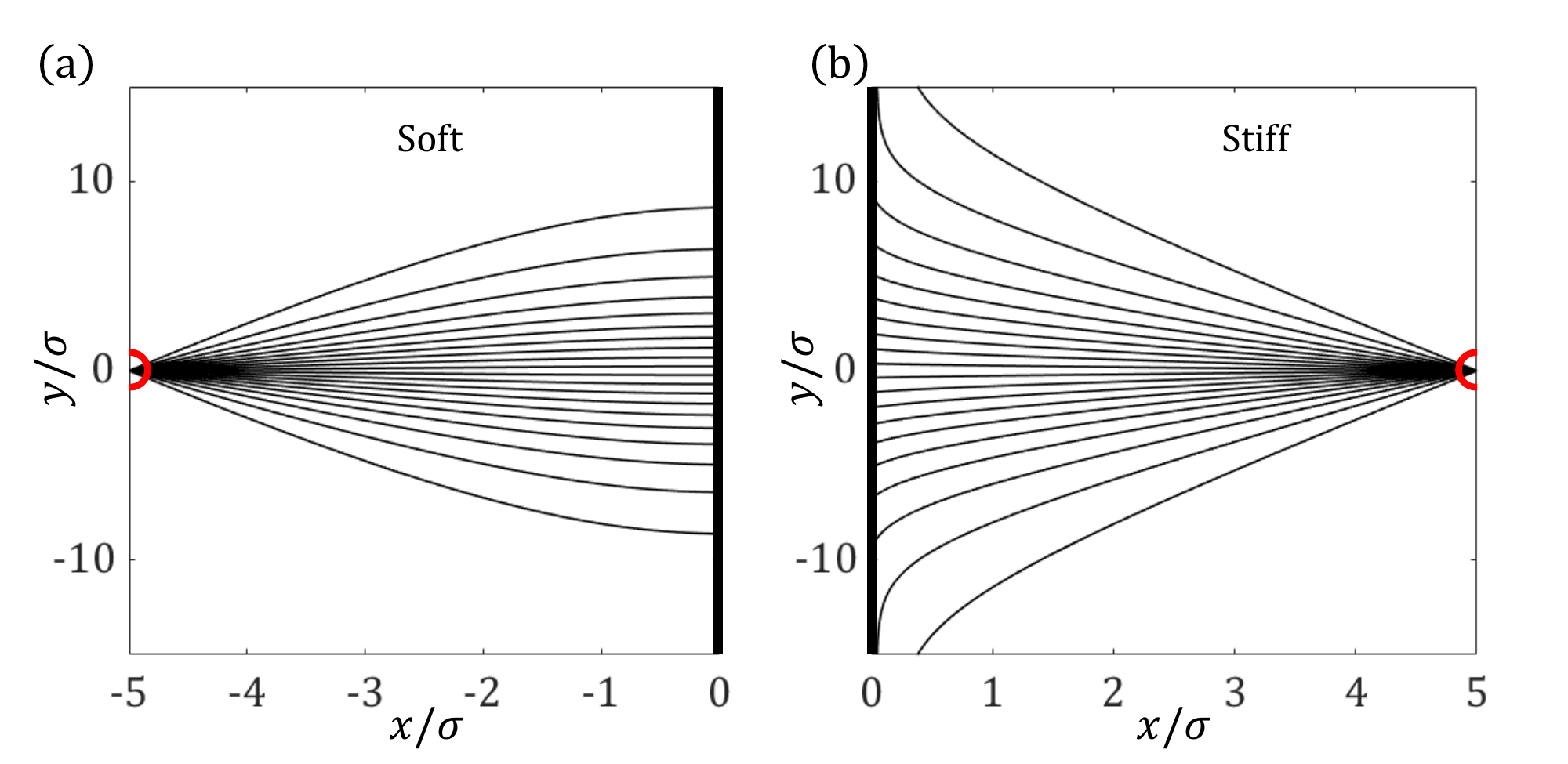}
\caption{The elastic torque from interface between the soft and stiff substrate tries to align the cells perpendicular or parallel to the boundary depending on the direction of approach to the interface. In these trajectories we do not consider any translational or rotational diffusion and also neglect the force from the interface. We observe the effect of elastic torque on the trajectory of cell approaching from the same distance on either side of the interface. (a) When cells are approaching from the softer side of the interface ($x = -5$, $y = 0$, shown by red semicircle) where the interface is at $x = 0$, with $Pe = 2$, the particle trajectories (shown by black lines) get aligned perpendicular to the interface. This kind of behavior is observed in case of refraction when light is traveling from a rarer to a denser medium. Here the refractive index can be realized to be increasing as it approaches the interface. The torque parameter $B = 5$ in the softer substrate region. (b) Cells approaching the interface from the stiffer side of the interface ($x = 5$, $y = 0$, shown by red semicircle) aligns parallel to the interface. This kind of of behavior is observed when light is traveling from a optically denser to a rarer medium. The torque parameter $B = 1$ in the softer substrate region.}
\label{fig:Refraction}
\end{figure*}

Here we summarize the methodology used to compute the various tactic indices used in the main text, from our simulations. One of these metrics, the FMI is used to compare our results with the analysis of experiments in Ref.~\cite{duchez2019durotaxis}. In these experiments, cells were tracked every 15 minutes and the FMI was calculated at $1$ hour intervals, for a total period of $24$ hours.  The authors also quantified the persistence of trajectories by estimating the ratio of displacement and the distance covered by the cell in these $1$ hour intervals.  This quantity was measured to be approximately $0.35$ for cells moving in all three regions. The persistence time -- that is, the time over which the cells travel more-or-less in the same direction -- is around $0.1$ hour.  To calculate the FMI defined in Ref.~\cite{duchez2019durotaxis} from our simulations, the following procedure was used. The position of a test cell is tracked every ${\Delta}T = 2.5$ dimensionless times ($15$ minutes in experiments). From these positions, the FMI is evaluated every ${\Delta}T = 10$ times ($1$ hour in experiments). Cells move in the domain and sometimes upon reaching the boundary move along it. We do not consider the contribution of particles trapped at the boundary or traveling along the boundary in the FMI calculation, since the biological experiments are in unconfined geometries (corresponding to cells at the boundary just crossing over). We also evaluated the FMI for only up to a dimensionless observation time $T = 10$, since most particles reach the boundary within that time. We calculate the FMI inside a region $6\sigma$ from the boundary, since we already established that beyond this region the influence of the elastic potential of the boundary is very low.

 To calculate the durotactic index (DI), we combine the results of two different simulations, $A\neq0$ and $A=0$, corresponding to the same value of $Pe$. Here, $A$ ranges from $0 - 10$. In these simulations, we consider the same elastic force and torque interaction parameters for cells with the boundary such that $A=B$. Then, we consider the difference in the steady state numbers of cells localized at the boundary between $A\neq0$ and $A=0$, to calculate the DI. 
 We observe that DI increases with $A$ and reaches a limiting value, for all values of $Pe$ (Fig.~\ref{fig:DI_vs_A_and_Pe}(a)), while for given $A$, the DI decreases with $Pe$ (Fig.~\ref{fig:DI_vs_A_and_Pe}(b)).

\section{Scattering of motile cells across elastic interface}
 
Here we analyze the deterministic trajectories (without noise) of cells across a sharp gradient of substrate stiffness. The interface is located at $x = 0$, where the $x<0$ is soft, with Young's modulus $5$kPa (and correspond to $A = B = 5$) and $x>0$ being stiff with Young's modulus $25$kPa (and correspond to $A = B = 1$). We consider clamped and free boundary conditions on the interface if the cells are on the soft and stiff surfaces respectively. The interaction potential obtained for substrate mediated elastic cell-boundary interaction is mentioned in App. A. We do not consider any rotational or translational diffusion for these trajectories. After initializing cells on either side of the interface, i.e. at $x = -5$ and $x = 5$, at different angles (20 angles linearly distributed between $-1$ to $1$ radian when on soft side of the substrate and $\pi-1$ to $\pi+1$ when initialized on stiff side of the substrate), we observe how the orientation of the cells change as they approach the interface. 

\begin{figure*}[htbp!]
\centering
\includegraphics[width=0.95\linewidth]{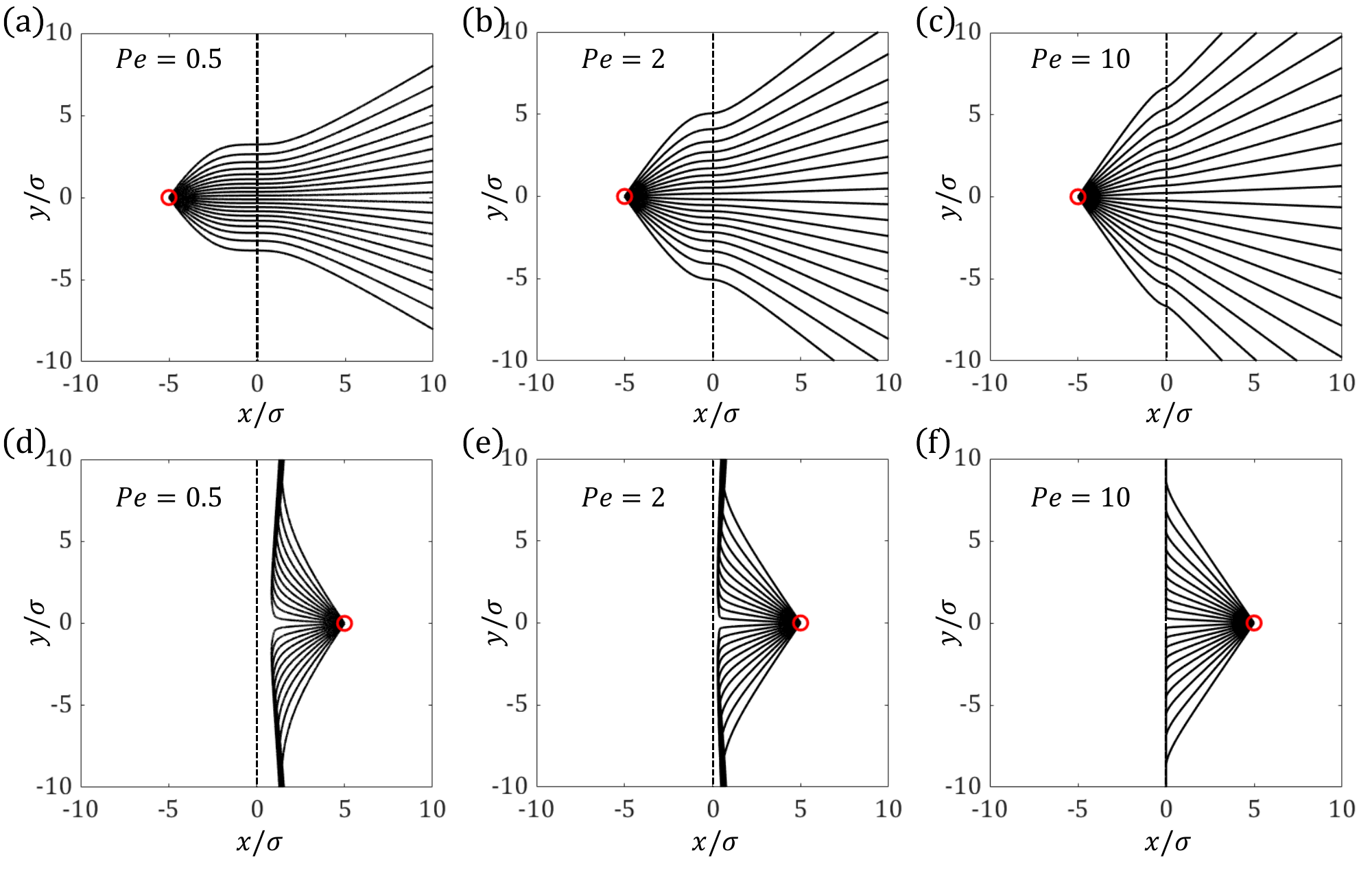}
\caption{The trajectories that are allowed to cross the interface are shown by black lines. The cells approaching the interface (shown by dashed line at $x = 0$) from either side of the interface have motility (a,d) $Pe = 0.5$, (b,e)$Pe = 2$ and (c,f)$Pe = 10$. The soft side of the substrate ($x<0$) has a Young's modulus of 5kPa ($A = B = 5$), while the stiff side ($x>0$) of the substrate has a Young's modulus of 25 kPa ($A = B = 1$).}
\label{fig:Refraction_2}
\end{figure*}
First, we focus on a simple condition where force from the interface is $0$, but cells are still acted on by the elastic torque from the substrate interface, i.e. $A = 0$ (but $B \neq 0$). We obtain trajectories of cells approaching the substrate interface and have a motility of $Pe = 2$ (Fig.~\ref{fig:Refraction}). The cells initialized on the soft side of the substrate prefer to align perpendicular to the interface when they approach it (Fig.~\ref{fig:Refraction}a). This behavior resembles light refraction when it moves from a rarer to a denser medium. The gentle transition of incident angle suggests a gentle increase in the refractive index of the medium. On the other hand, cells initialized on the stiffer side of the substrate prefer to bend parallel to the interface while approaching (Fig.~\ref{fig:Refraction}b). This behavior resembles light refraction from a denser to a rarer medium. Decreasing P{\'e}clet number implies slower moving cells that get more time to reorient. This is analogous to higher contrast of refractive index as light approaches closer to the interface, while for higher P{\'e}clet number, the contrast in refractive index is smaller.

If we consider the non-zero force from the boundary, that is, $A = B$, cells moving from the soft side to the stiffer side cross over (Fig.~\ref{fig:Refraction_2}a,b,c) even with low motility ($Pe = 0.5, 2$, shown in Fig.~\ref{fig:Refraction_2}a,b), because of the attractive nature of the clamped boundary condition and they reorient perpendicular to the interface when they cross over. After they have crossed, the elastic torque scatters their trajectories. 
At high motility, $Pe = 10$ (Fig.~\ref{fig:Refraction_2}c), cells do not have enough to sense torque and do not scatter much from the initial angle of incidence. On the other hand, cells approaching from the stiff side of the substrate (Fig.~\ref{fig:Refraction_2}d,e,f) do not cross the interface at low motility ($Pe = 0.5$ and $2$, as shown in Fig.~\ref{fig:Refraction_2}d,e respectively) as they are pushed away by the repulsive free boundary. At high motility ($Pe = 10$, as shown in Fig.~\ref{fig:Refraction_2}f) although the cells manage to reach the interface, the strong clamped boundary condition with interaction strength $A = 5$ on the soft side of the substrate does not allow the cells to move any further and crawl along the interface.
\bibliography{biblio}

\end{document}